\documentclass[prc,aps,nofootinbib,superscriptaddress,showkeys,showpacs,twocolumn,10pt,floatfix]{revtex4-2}
\usepackage{epsfig}
\usepackage{graphicx}
\usepackage[small]{subfigure}
\usepackage{setspace}
\usepackage[T1]{fontenc}
\usepackage[utf8]{inputenc}
\usepackage[english]{babel}
\usepackage{mathrsfs}
\usepackage{braket}
\usepackage[overload]{empheq} 
\usepackage{amssymb}
\usepackage{amsmath}
\usepackage{amsfonts}
\usepackage{siunitx}
\usepackage{amsopn}
\usepackage{mathtools}
\usepackage{float}
\usepackage{chemformula}
\usepackage{comment} 
\usepackage{tikz}
\usepackage{booktabs}
\usepackage{multirow}
\usepackage{verbatim}
\usepackage{xcolor}
\usepackage{ulem}
\usepackage{soul}
\usetikzlibrary{arrows}
\usepackage[colorlinks,linkcolor=blue,urlcolor=blue,citecolor=blue]{hyperref}

\newcommand{\blu}[1]{\textcolor[rgb]{0.0,0.0,1.0}{\bf (comment: #1)}}

\newcounter{singpart}[subsection]
\renewcommand{\thesingpart}{\thesubsection.\arabic{singpart}}

\newcommand{\singpart}[1]{%
  \par\addvspace{0.8\baselineskip}%
  \refstepcounter{singpart}%
  \noindent\textbf{\thesingpart\quad #1}\par\addvspace{0.3\baselineskip}%
}
\renewcommand{\sout}{\bgroup \color{red} \ULdepth=-.5ex \ULset}

\definecolor{lime}{HTML}{A6CE39}
\DeclareRobustCommand{\orcidicon}{
	\begin{tikzpicture}
	\draw[lime, fill=lime] (0,0) 
	circle [radius=0.16] 
	node[white] {{\fontfamily{qag}\selectfont \tiny ID}};
	\draw[white, fill=white] (-0.0625,0.095) 
	circle [radius=0.007];
	\end{tikzpicture}
	\hspace{-2mm}
}

\foreach \x in {A, ..., Z}{%
	\expandafter\xdef\csname orcid\x\endcsname{\noexpand\href{https://orcid.org/\csname orcidauthor\x\endcsname}{\noexpand\orcidicon}}
}

\begin{document}

\title{Spinodal instability in nuclear matter with light cluster degrees of freedom}

\author{Stefano Burrello\orcidA{}}
\email[]{burrello@lns.infn.it}
\affiliation{INFN, Laboratori Nazionali del Sud, I-95123 Catania, Italy}
\author{Carmelo Piazza}
\email[]{carmelo.piazza@phd.unict.it}
\affiliation{Department of Physics and Astronomy “Ettore Majorana”, University of Catania,
Via Santa Sofia 64, I-95123 Catania, Italy}
\affiliation{INFN, Laboratori Nazionali del Sud, I-95123 Catania, Italy}
\author{Rui Wang\orcidC{}}
\email[]{rui.wang@ct.infn.it}
\affiliation{INFN, Laboratori Nazionali del Sud, I-95123 Catania, Italy}
\author{Maria Colonna\orcidB}
\email[]{colonna@lns.infn.it}
\affiliation{INFN, Laboratori Nazionali del Sud, I-95123 Catania, Italy}

\date{\today}

\begin{abstract}
We investigate the thermodynamical stability of low-density isospin-symmetric
nuclear matter at finite temperature, explicitly including light clusters as
degrees of freedom. Within a generalized mean-field framework, we compute the
curvature matrix of the free-energy density and determine the spinodal region,
identifying the conditions under which mechanically unstable modes may develop
in the presence of clustering. Particular attention is devoted to the formal
consequences of introducing an infrared momentum cutoff in the density and
current moments, which effectively accounts for Pauli-blocking effects and the
associated reduction of low-momentum quasiparticle states in the medium. We
show that when the cutoff is density dependent, thermodynamic consistency
requires additional contributions to the chemical potentials
and extra terms also appear in the first hydrodynamic moment,
influencing both the
stability analysis and the location of the spinodal boundary. We further
examine the character of the unstable modes and find that a sufficiently stiff
density dependence of the cutoff may drive clusters to fluctuate out of phase
with nucleons, pushing them toward low-density regions while nucleonic
instabilities grow, in contrast with the in-phase pattern obtained when the density derivatives of the cutoff are neglected in the stability analysis. Our results shed new light on the role of light clusters in the phase dynamics of warm, dilute nuclear matter, with
implications for heavy-ion collisions and for the physics of neutron-star
crusts.
\end{abstract}

\maketitle

\section{Introduction}

The study of nuclear matter under extreme conditions remains a central theme in
nuclear physics, with profound implications for compact stellar objects and dense astrophysical environments~\cite{lattimerPREP2016, oertelRMP2017, BurgioPPNP2021}. A key aspect of this investigation lies in
understanding the thermodynamic properties and composition of nuclear matter
across a wide range of densities and temperatures, and therefore on its equation of state (EoS)~\cite{HorowitzNPA2006, radutaPRC2010, typelPRC2010, burrelloEPJA2022, KellerPRL2023}. In particular, the
sub-saturation regime hosts a rich and complex phenomenology shaped by both
mean-field effects and many-body correlations~\cite{riosPRC2014, henRMP2017, burEPJA2022}.

At these low densities, nucleons do not simply behave as a uniform Fermi gas;
instead, they tend to form bound states, including light clusters such as
deuterons, tritons, and $\alpha$ particles~\cite{paisPRC2019, WangArXiV2025}. These few-body correlations emerge
as a natural way for the system to minimize its free energy and dominate the
composition at sufficiently low densities and moderate temperatures~\cite{typelPRC2010, burrelloPRC2015, WuJLTP2017}. However,
as the density increases, medium modifications---most notably Pauli
blocking effects---progressively reduce the binding energies of these clusters,
eventually leading to their dissolution in what is known as the Mott transition~\cite{ropkePRC2009, ropkeNPA2011}.

At sub-saturation densities, nuclear matter may also become unstable against
long-wavelength density fluctuations~\cite{BaranNPA1998}. When the incompressibility turns
negative, the mean-field attraction drives the system into the spinodal region of the liquid--gas phase diagram~\cite{ChomazPR2004}, where mechanical instabilities amplify density fluctuations, possibly leading to phase separation and the formation of inhomogeneous structures~\cite{baranPREP2005}.

However, light cluster formation 
is expected to modify both the thermodynamics and the dynamical response, introducing additional modes associated with cluster densities and currents~\cite{KanadaPTEP2012}. As a consequence, capturing the full complexity of low-density nuclear matter, especially in out-of-equilibrium conditions, remains a major challenge~\cite{onoPPNP2019}. 

Thermodynamical approaches that include clustering effects are indeed often limited to equilibrium properties~\cite{RoepkePRC2020} or do not take explicitly into account the role of medium modifications of light clusters to shape the onset of instabilities~\cite{AlamPRC2017, BoukariPRC2021}. On the other hand, traditional dynamical models, such as those based on transport approaches~\cite{TMEP, XuPPNP2019}, while successful in describing nucleonic
dynamics and collective flow~\cite{NapolitaniPLB2013, LinPRC2018}, often struggle to treat light clusters as
correlated, dynamical degrees of freedom on the same footing as nucleons. 
Therefore, a unified framework connecting clustering, in-medium effects, and mechanical instabilities is still lacking, despite recent significant progresses~\cite{DanNPA533, WangPRC2023,chengPRC2024}. 

Clarifying how mean-field instabilities develop in the presence of light clusters is crucial for applications in both nuclear physics
and astrophysics~\cite{PaisPRC2015}. 
On the terrestrial side, heavy-ion collisions at
Fermi and intermediate energies probe nuclear matter far from equilibrium~\cite{liPREP2008, ColonnaPPNP2020}, providing crucial insights on the EoS~\cite{SorensenPPNP2024, TsangNAT2024}. Observables such as multifragmentation patterns,
cluster yields, and correlation functions are highly sensitive to the instability properties of dilute matter and to the role played by light clusters during the expansion phase~\cite{ onoPPNP2019, FrosinPRC2023, FablePRC2024,  chengPRC2024}. In astrophysical environments, including
core-collapse supernovae, neutron-star crusts, and binary neutron-star mergers,
the same low-density and moderate-temperature conditions govern the formation
of nonuniform structures and influence neutrino transport, thermal evolution,
and potentially the emission of gravitational waves~\cite{SumiyoshiPRC2008,   ArconesPRC2008, burrelloPRC2016, YudinMNRAS2018, FischerPRC2020, DavisAA2024, NgAJ2024, RadutaAA2025, BurrelloPRC2025}.

Specifically, spinodal decomposition must be viewed as a coupled response of nucleons and clusters embedded in a self-consistent medium. In this framework, in-medium effects acting on clusters may be implemented through an infrared density-dependent Mott momentum~\cite{ropkePRC2015}, which effectively accounts for the Pauli-blocking reduction of low-momentum quasi-particle states in the medium. Within this assumption,  linear response studies of the Vlasov dynamics have recently shown that in-medium effects can significantly reshape the unstable response of dilute nuclear matter~\cite{WangPRC2024}. In particular, the density dependence of the cutoff may alter the instability growth and even modify the relative pattern of density fluctuations between clusters and nucleons, with respect to calculations neglecting the density derivatives of the cutoff in the stability analysis. 
However, the features of these instability patterns, particularly in connection with clustering and in-medium effects, require further study. In this work, we explore whether new insights can be obtained within a genuine thermodynamic analysis 
of the curvature of the free-energy density, so without resorting to dynamical prescriptions. We 
address this issue by investigating 
 the thermodynamic stability of low-density nuclear matter at finite temperature, explicitly including light clusters. Within a generalized mean-field framework, 
we assess how cluster formation and in-medium effects modify the spinodal region, with particular emphasis on the role of the infrared momentum cutoff. 
Building on earlier studies limited to a single cluster species, we now allow for the simultaneous inclusion of multiple species 
and explore the sensitivity of the stability properties to different assumptions on cutoff parameterizations, nucleon-cluster mean-field potential strength and in-medium binding energies. 
When the cutoff depends on density, thermodynamic consistency requires additional contributions to the chemical potentials, whose impact on free-energy convexity and unstable modes is systematically examined.
We further analyze the directions of unstable fluctuations, to characterize the nature of the 
normal modes and determine whether clusters cooperate with or decouple from nucleonic density fluctuations. The present study thus complements the recent linear-response analysis of Ref.~\cite{WangPRC2024} by providing a more extensive 
perspective on the role of clustering in the behavior of warm and dilute nuclear matter.

The paper is organized as it follows: Sec.~\ref{sec:theo} outlines
the theoretical framework and presents the connection 
with hydrodynamics and the Vlasov approach at the onset of the spinodal instability, Sec.~\ref{sec:results} is dedicated to the results and conclusions are drawn in Sec.~\ref{sec:conclusions}.

\section{Theoretical framework}
\label{sec:theo}

\subsection{Distribution functions and Mott momentum}

We consider a system of nucleons, neutrons ($n$) and protons ($p$), and light nuclear clusters treated as explicit degrees of freedom, in thermodynamical equilibrium at temperature $T$. The equilibrium phase-space distribution functions $f_j$ are given by the Fermi--Dirac or Bose--Einstein form
\begin{equation}
f_{j}(\epsilon_{j}) =
\left[
\exp\!\left(\dfrac{\epsilon_{j}-\mu_{j}^\ast}{T}\right)
-(-1)^{A_{j}}
\right]^{-1},
\label{eq:distribution_function}
\end{equation}
where $A_j$ is the mass number of species $j$ and $\mu_j^\ast$ denotes the nonrelativistic effective chemical potential. 
The single-particle kinetic energy is
\begin{equation}
\epsilon_{j} = \dfrac{p^{2}}{2m_{j}},
\end{equation}
with the constituent mass defined as
\begin{equation}
m_{j} = A_{j}m - B_{j},
\end{equation}
where $B_j$ is the absolute value of the binding energy of species $j$ and $m=939~\mathrm{MeV}$ is the bare nucleon mass, taken to be the same for neutrons and protons.

The number density $\rho_j$ of each species is obtained from the momentum integral of the distribution function,
\begin{equation}
\rho_{j}
=
g_{j}
\int_{\Lambda_{j}}
\dfrac{d\mathbf{p}}{(2\pi\hbar)^{3}}
\, f_{j},
\label{eq:rhoi}
\end{equation}
where $g_j$ is the spin-degeneracy factor and the total baryon density is $\rho_{b} = \sum_{j} A_{j} \rho_{j}$. For nuclear clusters, an infrared momentum cutoff $\Lambda_j$ is introduced to account for the Mott effect. Indeed, due to Pauli blocking in the medium, bound clusters can exist only for momenta exceeding the corresponding Mott momentum, leading to a suppression of low-momentum quasiparticle states~\cite{ropkePRC2015, RoepkePRC2020, WangPRC2023, WangPRC2024}.

The momentum cutoff effectively incorporates medium effects analogous to a momentum-dependent binding-energy shift within a quasiparticle picture~\cite{ropkeNPA2011}. In the present approach, where a momentum cutoff is introduced in the
cluster distribution functions, surviving clusters are assumed to
retain their vacuum binding energies $B_{j}^{\rm vac}$, independently of
the surrounding nuclear medium, unless otherwise specified.
Correlations in the continuum, which may become important at higher densities~\cite{ropkePRC2015, burEPJA2022}, are also neglected.

\subsection{Free-energy density functional}
The thermodynamical properties of such a system might be completely determined through the study of its thermodynamical potential. In the finite temperature case we are focusing on, this potential is given by the free energy density functional $\mathcal{F} = \mathcal{E} - T \mathcal{S}$, where $\mathcal{S}$ is the entropy density given by
\begin{equation}
\mathcal{S} = - \sum_{j} g_{j} \int_{\Lambda_{j}} \dfrac{d \mathbf{p}}{(2\pi\hbar)^{3}} \left[ f_{j} \ln f_{j} + \dfrac{1 - \sigma_{j} f_{j}}{\sigma_{j}} \ln \left ( 1 - \sigma_{j} f_{j} \right) \right]
\end{equation}
with $\sigma_{j} =  (-1)^{A_{j}+1}$ 
and $\mathcal{E}$ is the energy density functional $\mathcal{E} = \mathcal{K} + \mathcal{U}$, expressed as the sum of the kinetic energy density 
\begin{equation}
\mathcal{K}  =  
\sum_{j} g_{j} \int_{\Lambda_{j}} \dfrac{d \mathbf{p}}{(2\pi\hbar)^{3}} 
f_{j} \epsilon_{j} 
\end{equation}
and the potential energy density $\mathcal{U}$, as obtained from a (density-dependent) effective interaction.
The free-energy density functional may be then finally written as
\begin{eqnarray}
\mathcal{F} &=&  \mathcal{U} + \sum_{j} \left(\mu_{j}^{\ast} + m_{j} \right) \rho_{j} \nonumber \\
& & -  T  
\sum_{j} \dfrac{g_{j}}{\sigma_{i }} \int_{\Lambda_{j}} \dfrac{d \mathbf{p}}{(2\pi\hbar)^{3}} \ln \left[ 1 + \sigma_{j} \exp\left( - \dfrac{\epsilon_{j} - \mu_{j}^{\ast}}{T}\right) \right]. \nonumber \\
\label{eq:functional}
\end{eqnarray}
In this work we concentrate on the temperature domain 
$T \gtrsim$ 5 MeV, which is relevant for the expansion phase of heavy-ion collisions at Fermi/intermediate energies (in the range of beam energies
$E /A \approx 30-300$ MeV/nucleon)~\cite{DengPRC2016} as well as to 
the temperature conditions 
created in binary neutron-star mergers and supernova explosions~\cite{SumiyoshiPRC2008, MostPRD2023}.
These temperature values mainly lie beyond the critical one for the transition to the Bose-Einstein condensate phase for the light clusters, whose emergence is further suppressed when introducing a momentum cutoff~\cite{WuJLTP2017}. Then 
the functional $\mathcal{F}$ does not include any contribution from boson condensation.

\subsection{Thermodynamical consistency and chemical potentials}

Equation~\eqref{eq:functional} extends the standard definition of the thermodynamical potential to the case in which an infrared momentum cutoff, either constant or density dependent, is introduced in the momentum integrals. 
Chemical potentials are consistently obtained from the thermodynamical potential as 
\begin{equation}
\mu_{j} = \left. \dfrac{\partial \mathcal{F}}{\partial \rho_{j}} \right|_{\{\rho_{l},\, l\neq j\}}.
\label{eq:mu1}
\end{equation}
As shown in Appendix~\ref{app:mu}, from Eq.~\eqref{eq:mu1}, the following relation is obtained
\begin{equation}
\mu_{j} = m_{j} + \mu_{j}^{\ast} + U_{j} +\tilde{\mu}_{j},
\label{eq:mu}
\end{equation}
where $\mu_j^\ast$ is the effective chemical potential appearing in the equilibrium distribution functions,  
\begin{equation}
U_{j} = \left.
\dfrac{\partial \mathcal{U}}{\partial \rho_{j}}
\right|_{\{\rho_{l},\, l\neq j\}},
\label{eq:potential}
\end{equation}
is the mean-field potential and $\tilde{\mu}_{j}$ are additional contributions 
arising from the density-dependent cutoff, which explicitly write 
\begin{equation}
\tilde{\mu}_{j}
= - \sum_{c} \sigma_{c} \alpha_{c} \sqrt{\lambda_{c}}
\left( \lambda_{c} - \mu_{c}^{\ast} + T \ln f_{c}^{\lambda}
\right) \left.
\dfrac{\partial \lambda_{c}}{\partial \rho_{j}}
\right|_{\{\rho_{l},\,l\neq j\}},
\label{eq:mutilde}
\end{equation}
where $c$ runs only over cluster species,
\begin{equation}
\alpha_{c} = g_{c} \dfrac{(2m_{c})^{3/2}}{4\pi^{2}\hbar^{3}} 
\label{eq:alpha}
\end{equation}
and we introduced the cluster kinetic energy cutoff $\lambda_c = \frac{\Lambda_c^2}{2m_c}$ and the shorthand notation $f_{c}^{\lambda}\equiv f_{c}(\lambda_{c})$. In the following, all partial derivatives with respect to $\rho_j$ are evaluated at fixed densities of the other species $\rho_l$ with $l\neq j$.
The appearance of the additional contribution to the chemical potential, $\tilde{\mu}_{j}$,
represents a genuine medium-induced rearrangement effect. 
An alternative strategy to enforce thermodynamical consistency would, in principle, consist in introducing 
an explicit rearrangement contribution $\mathcal{F}^{(r)}$ to the free-energy density functional 
\begin{equation}
\bar{\mathcal{F}} = \mathcal{F}
-  \mathcal{F}^{(r)},
\end{equation}
such that
\begin{equation}
\tilde{\mu}_{j} = \dfrac{\partial \mathcal{F}^{(r)}}{\partial \rho_{j}}
\label{eq:rearrangement}    
\end{equation}
and $\mu_j = m_{j} + \mu^\ast_j+ U_j$. 
However, the explicit construction of a rearrangement functional
$\mathcal{F}^{(r)}$ capable of reproducing Eq.~\eqref{eq:rearrangement} is not
straightforward in the most general case. 

Expanding
$\mathcal{F}$ up to second order in the density fluctuations yields
\begin{equation}
\label{eq:deltaF}
\delta \mathcal{F} - \sum_{j}\mu_{j}\,\delta\rho_{j}
= \frac{1}{2} \sum_{j,l}
\frac{\partial \mu_{j}}{\partial \rho_{l}}
\,
\delta\rho_{j}\,\delta\rho_{l},
\end{equation}
which involves the derivatives of the chemical potentials defined in
Eq.~\eqref{eq:mu}. These derivatives are explicitly derived in Appendix~\ref{app:dmu}. The thermodynamic stability of the system is therefore
controlled by the curvature matrix associated with the quadratic form in
Eq.~\eqref{eq:deltaF}, whose elements are
\begin{equation}
\mathbb{C}_{jl} = \frac{\partial \mu_{j}}{\partial \rho_{l}} .
\label{eq:c3}
\end{equation}
It is worth noticing that, since 
$\mathbb{C}$ is associated with a symmetric bilinear form, it must be symmetric by construction, even in the presence of a density-dependent infrared cutoff. The eigenvectors of $\mathbb{C}$ 
define the normal modes in the space of the density fluctuations, while the associated eigenvalues determine their stability: positive eigenvalues
correspond to stable modes, whereas negative eigenvalues signal the onset of
density instabilities.

\subsection{Linearized Vlasov equations}

In Ref.~\cite{WangPRC2024}, an extended dynamical framework was introduced to describe the development of volume instabilities in the presence of light clusters. The approach is based on a linear response analysis of the collisionless (Vlasov) limit of the Boltzmann equation~\cite{ColonnaPRC2008, burrelloPRC2019, ShvedovPRC2025}. Following this strategy, we consider phase-space small-amplitude perturbations $\delta f_j$ around the equilibrium distribution functions $f_j$. The resulting linearized Vlasov equations read
\begin{equation}
\label{eq:vlasov_linear}
\frac{\partial (\delta f_{j})}{\partial t}
+
\nabla_{\mathbf{r}} (\delta f_{j}) \cdot \nabla_{\mathbf{p}} \varepsilon_{j}
-
\nabla_{\mathbf{p}} f_{j} \cdot \nabla_{\mathbf{r}} (\delta \varepsilon_{j})
=
0,
\end{equation}
where $\varepsilon_j$ is the single-particle energy and $\delta \varepsilon_j$ its fluctuation (see Supplemental Materials of Ref.~\cite{WangPRC2024}).

The single-particle energy is related to the functional derivative of the energy density according to
\begin{equation}
\varepsilon_{j}
\equiv
\dfrac{1}{\alpha_{j}\sqrt{\epsilon_{j}}}
\dfrac{\delta \mathcal{E}}{\delta f_{j}},
\end{equation}
with $\alpha_{j}$ defined as in Eq.~\eqref{eq:alpha}.
In the presence of a density-dependent momentum cutoff, the single-particle energy can be written as
\begin{equation}
\varepsilon_{j} = \epsilon_{j}
+ U_{j} + \tilde{\varepsilon}_{j},
\end{equation}
where 
\begin{equation}
\tilde{\varepsilon}_{j} = - \sum_{c} \dfrac{\lambda_{c}+U_{c}}{1+\Phi^{cc}}
\,
\Phi^{cj},
\label{eq:epsilonlambda}
\end{equation}
with
\begin{equation}
\Phi^{cj} = \alpha_{c}\sqrt{\lambda_{c}}\,
f_{c}^{\lambda}
\dfrac{\partial \lambda_{c}}{\partial \rho_{j}},
\label{eq:Phi}
\end{equation}
denotes the additional contribution induced by the density dependence of the cutoff. 

For the momentum-independent effective interaction considered in this work, the derivation of the linearized Vlasov equations has been presented in Ref.~\cite{WangPRC2024}. Extending the standard Landau procedure to include multiple cluster species $c$, one obtains a system of coupled equations for the density fluctuations that can be written in the compact form
\begin{equation}
\delta \rho_{j}
=
-
\chi_{j}
\sum_{l}
\left(
F_{0}^{jl}
+
\tilde{F}^{jl}
\right)
\delta \rho_{l}
-
\sum_{c}
\delta_{jc}
\sum_{l}
\Phi^{cl}
\delta \rho_{l},
\label{eq:deltarhoj}
\end{equation}
where $\chi_j$ is the 
Lindhard function and $l$ runs over all species. The standard Landau parameters are defined as
\begin{equation}
F_{0}^{jl}
=
N_{j}
\dfrac{\partial U_{j}}{\partial \rho_{l}},
\label{eq:landau}
\end{equation}
and, in analogy, the additional contributions associated with the density-dependent cutoff read
\begin{equation}
\tilde{F}^{jl} =
N_{j} \dfrac{\partial \tilde{\varepsilon}_{j}}{\partial \rho_{l}},
\label{eq:landau_lambda}
\end{equation}
where
\begin{equation}
N_{j}
=
-
g_{j}
\int_{\Lambda_{j}}
\dfrac{d\mathbf{p}}{(2\pi\hbar)^{3}}
\dfrac{\partial f_{j}}{\partial \epsilon_{j}}
\label{eq:nj}
\end{equation}
is the thermally averaged level density of species $j$.

\subsection{Euler equations and hydrodynamics}
\label{sec:hydro}

The hydrodynamical approach is usually adopted to describe the long-wavelength
dynamics of a classical fluid dominated by
(quasi-)particle binary collisions. In this regime, 
the 
two-body correlations drive 
the system toward local thermodynamic equilibrium on a time scale that is assumed to be short
compared with that associated with the collective evolution~\cite{ChomazPR2004,
zhengPLB2017}. 
In the ideal limit, local thermodynamical equilibrium is assumed at each time instant and dissipative effects, such as viscosity and heat conduction, are  neglected. 
As a consequence, although hydrodynamics is intrinsically a collisional regime,
the resulting ideal hydrodynamic equations coincide with the lowest moments of
the collisionless (Vlasov) equation once local equilibrium is 
enforced. 
Accordingly, the zeroth (continuity) and first moment (Euler) equations can be formally recovered from
the moment expansion of the Boltzmann transport equation~\cite{ChomazPR2004, PaisPRC2015}.
The detailed derivation is presented in
Appendix~\ref{app:euler}, where the case of momentum integrals including a density-dependent infrared cutoff is explicitly addressed. The resulting equations take the form
\begin{eqnarray}
\frac{\partial \rho_{j}}{\partial t} + \nabla_{\mathbf r}\!\cdot\! \left(
\rho_{j}\mathbf{u}_{j} \right)
&=& 0,
\nonumber\\[1ex]
m_{j}^{\lambda}\,
\frac{\partial}{\partial t}
\left( \rho_{j}\mathbf{u}_{j}
\right) + \nabla_{\mathbf r} P
+ \nabla_{\mathbf r}\tilde{P} &=& 0,
\label{eq:euler}
\end{eqnarray}
where $\mathbf{u}_{j}$
is the average velocity (see Eq.~\eqref{eq:avg_velocity}) 
and the other quantities are described below. 

Equations~\eqref{eq:euler} 
differ from the standard hydrodynamic equations through two distinct mechanisms
associated with the infrared cutoff $\lambda_{j}$. On the one hand, even when the cutoff is constant, the suppression of the
momentum region $|\mathbf{p}|<\Lambda_{j}$ reduces the available phase space and
shifts the distribution toward higher momenta. Physically, this suppresses soft
(low-momentum) quasiparticle excitations, thereby modifying the inertial
response of the fluid. This 
leads to 
the renormalized inertial mass 
\begin{equation}
m_{j}^{\lambda}
=
\frac{m_{j}}{\mathcal{Z}_{j}^{\lambda}},
\end{equation}
with the factor
\begin{equation}
\mathcal{Z}_{j}^{\lambda}
=
1
+
\frac{\tfrac{2}{3}\alpha_{j} f_{j}^{\lambda}\lambda_{j}^{3/2}}{\rho_{j}}.
\label{eq:zeta}
\end{equation}
The resulting picture is that of an effective quasiparticle ensemble in which
only modes above the threshold $\lambda_{j}$ participate in transport, while
the low-energy sector is frozen out. On the other hand, if the cutoff $\lambda_{j}$ depends on the local densities $\rho_{l}$, further modifications arise. In this case, the integration boundary in momentum space varies in space and time, and the derivatives of $\lambda_{j}$ generate an additional 
pressure contribution, $\tilde{P}$, 
in the momentum equation, 
whereas the pressure $P$ retains its standard functional form: 
\begin{equation}
P = \sum_{j}\mu_{j}\rho_{j} -
\mathcal{F}.
\label{eq:pressure}
\end{equation}

Within the hydrodynamic description, the stability and evolution of density
fluctuations are finally governed by  the derivative, with respect to the densities of the different species, of the term ($P+\tilde{P}$). 
If the derivatives of $\tilde{P}$ are small compared with the standard compressional response, $\partial P/\partial \rho_{l}$—as is typically the case for smooth cutoff parametrizations and moderate cluster abundances—the $\tilde{P}$ term can be regarded as a subleading correction. In this limit, 
the stability analysis 
reduces to the study of the curvature of the free-energy density functional $\mathcal{F}$~\cite{BaranPRL2001,burrelloPRC2014}. 

In the next section, we clarify the connection
between 
the analysis based on the curvature of the free-energy density and the linearized Vlasov approach, when a nonstandard
infrared momentum cutoff is adopted in the momentum integrals.

\subsection{Connection between  thermodynamical analysis and Vlasov approach}
\label{subsec:hydro_vlasov}
The connection between 
thermodynamical instabilities 
and the linearized Vlasov approach is widely investigated in the literature for the standard case~\cite{ChomazPR2004}. In that case, the stability condition is governed by the free-energy curvature matrix in both frameworks, which then provide two alternative ways of analyzing the extension of the spinodal region,  yielding identical predictions for the onset of the instability. However, when a density-dependent cutoff is introduced in the momentum integrals, the rearrangement contributions prevent this equivalence from holding in general, as already anticipated in the previous section. 

Let us examine this issue in more detail to clarify the previous statement.  For this purpose, let us consider the simplified case of symmetric nuclear matter (SNM) with only one isospin-symmetric cluster species $d$ added as explicit degree of freedom.  Under these conditions, the system is invariant under neutron-proton exchange, and it is convenient to introduce, together with the cluster density $\rho_{d}$ and the corresponding density fluctuation $\delta \rho_{d}$, the
combinations
\[
\delta\rho_{S,V} = \delta\rho_{n} \pm \delta\rho_{p},
\]
which describe, respectively, the fluctuations $\delta\rho_{S}$ of the
isoscalar density $\rho_{S}=\rho_{n}+\rho_{p}$ and the fluctuations
$\delta\rho_{V}$ of the isovector density $\rho_{V}=\rho_{n}-\rho_{p}$. 
It is worth noticing that, in the space of $(\delta \rho_{V}$,$\delta \rho_{S}$,$\delta \rho_{d})$ density fluctuations, the $3 \times 3$ matrix associated with the linear system of Eq.\ \eqref{eq:deltarhoj} 
acquires a block-diagonal structure, and the isovector sector ($\delta\rho_{n}=-\delta\rho_{p}$) decouples, since the corresponding eigenmode is unaffected by the presence of an isospin-symmetric cluster. The nontrivial coupling between nucleons and clusters is therefore entirely contained in the isoscalar--cluster block.  For $\chi_{q}=\chi_{d}=1$, introducing the Landau parameter
\begin{equation}
F_{0} = F_{0}^{qq} + F_{0}^{qq^{\prime}},
\qquad q,q^{\prime}=n,p ,
\end{equation}
the onset of the SNM spinodal instability is then finally obtained by imposing that the determinant of the isoscalar--cluster block
\begin{equation}
\mathbb{A}
=
\begin{pmatrix}
1 + F_{0} + 2\tilde{F}^{qq}  & 2F_{0}^{qd} + 2 \tilde{F}^{qd}  \\
F_{0}^{dq} + \tilde{F}^{dq} + \Phi^{dq}  & 1 + F_{0}^{dd} + \tilde{F}^{dd} + \Phi^{dd}
\end{pmatrix}
\label{eq:a2lambda_tilde}
\end{equation}
vanishes. It is worth noting that the same instability boundary would be obtained from the linearized Euler equations. Moreover, from the above matrix, one readily recovers both the case of a 
density-independent 
cutoff 
and the limit of pure nucleonic matter, whose spinodal border is given by the standard condition $1+F_{0}=0$. 

Similarly to the linearized Vlasov approach, in the study of thermodynamic stability through the free-energy curvature matrix one can safely restrict the analysis to the following $2\times2$ symmetric matrix (see the derivation in Appendix~\ref{app:dmu}): 
\begin{equation}
\mathbb{C} 
=
\begin{pmatrix}
\dfrac{1 + F_{0} + 2\tilde{\Phi}^{qq}}{2N_{q}} &
\dfrac{F_{0}^{qd} + \tilde{\Phi}^{qd}}{N_{q}} \\[1ex]
\dfrac{F_{0}^{dq} + \tilde{\Phi}^{dq} + \Phi^{dq} }{N_{d}} &
\dfrac{1 + F_{0}^{dd} + \tilde{\Phi}^{dd} + \Phi^{dd}}{N_{d}}
\end{pmatrix}\hspace*{-0.4em}, 
\label{eq:c2lambda}
\end{equation}
where we have 
introduced the terms 
\begin{equation}
\tilde{\Phi}^{jl} = N_{j}\,\frac{\partial \tilde{\mu}_{j}}{\partial \rho_{l}},
\label{eq:phi_tilde}
\end{equation}
that account for the density derivatives of the additional contributions to
the chemical potentials. 
It is worth noting that, in the general case when the density dependence of the cutoff is taken into account, the zeros of the determinant of
$\mathbb{C}$ do not coincide with those of the
matrix $\mathbb{A}$ defined in
Eq.~\eqref{eq:a2lambda_tilde}, since $\tilde{\Phi}^{jl}\neq
\tilde{F}^{jl}$.
However, in the limit where the difference $(\tilde{F}^{jl} - \tilde{\Phi}^{jl}) 
$, namely the derivatives of the $\tilde{P}$ contribution entering Eq.~\eqref{eq:euler} can be neglected, the two approaches yield the same results. This also happens if we consider the stronger assumption 
of neglecting the contributions $\tilde{\varepsilon}_{j}$ and $\tilde{\mu}_{j}$ in the single-particle energies and chemical potentials, respectively, i.e., setting $\tilde{\Phi}^{jl} = \tilde{F}^{jl}$ = 0 in Eqs.~\eqref{eq:a2lambda_tilde} and~\eqref{eq:c2lambda}.
In the following, similarly to Ref.~\cite{WangPRC2024}, we will refer to the latter case as a ``hybrid'' situation, where the effects of the density-dependent cutoff appear only through the $\Phi^{jl}$ terms. 

The next section is devoted to present the results of the thermodynamic stability analysis of the composed system of nucleons and light clusters through the study of the free-energy curvature matrix. 
To better isolate the role of in-medium effects, we consider, in addition to the full case, a simplified limit in which the density derivatives of the cutoff are neglected in the stability analysis ($\Phi^{dj} = 0$). This also implies $\tilde{F}^{dj} = \tilde{\Phi}^{dj} = 0$. We refer to this situation as ``neglecting in-medium effects'', although the cutoff is still retained in the momentum integrals and continues to affect, among other quantities, the cluster fractions. 
Within a generalized mean-field framework, we assess how the spinodal region is modified by the presence of light clusters and in-medium effects, also comparing the results with the linear-response analysis based on the Vlasov equations of Ref.~\cite{WangPRC2024}.

\section{Results}
\label{sec:results}

The results of the present analysis are restricted to the case of SNM. The extension to asymmetric nuclear matter will be addressed in a forthcoming work.

We adopt a simplified Skyrme-like effective interaction for the mean-field potential, following Ref.~\cite{baranPREP2005}. In addition to the standard assumption that nucleons bound in clusters experience the same mean-field potential as free nucleons, we also explore alternative prescriptions inspired by RMF studies~\cite{burEPJA2022}. These works indicate that in-medium effects may require a reduced nucleon--meson coupling strength for clusters, typically implemented through a universal screening factor smaller than unity~\cite{paisPRC2019}. Such a prescription has been shown to improve the description of the EoS in astrophysical applications~\cite{paisPRL2020} and of the chemical equilibrium features characterizing heavy-ion collisions~\cite{qinPRL2012}. Within the present non-relativistic Skyrme framework, this effect can be encoded by assuming that the potential energy density $\mathcal{U}(\rho_{\eta})$ depends on a ``screened'' density
\begin{equation}
\rho_{\eta} = \rho_{S} + \eta\sum_{c} A_{c}\,\rho_{c},
\label{eq:eta}
\end{equation}
where $\eta\leq 1$. 

Moreover, we restrict ourselves to isospin-symmetric light clusters, such as
deuterons and/or $\alpha$ particles. Then, following Ref.~\cite{WangPRC2024}, the in-medium effects 
are parametrized through the following expression for the
kinetic-energy cutoff:
\begin{equation}
\lambda_{c}(\rho_{b}, T)
=
\beta_{c}\, \rho_{b}^{\gamma_{c}} \,
S_{c}(\rho_{b}, T), \qquad c = d, \alpha
\label{eq:EMt}
\end{equation}
which assumes a power-law dependence on the total baryon density $\rho_{b}$
\cite{KuhrtsPRC2001, ropkeNPA2011, WangPRC2023}, modulated by the smoothing
function
\begin{equation}
S_{c}(\rho_{b}, T)
= 1 + \tanh\!\left[
\xi_{c} \left( 1 - \nu_{c}
\frac{\rho_{c}^{\mathrm{Mott}}(T)}{\rho_{b}}
\right)
\right].
\label{eq:smoothing}
\end{equation}
The parameters $\beta_{c}$, $\gamma_{c}$, $\xi_{c}$, and $\nu_{c}$ control the
density dependence and smoothness of the cutoff. The function $S_{c}$ is
introduced to prevent the appearance of discontinuities in the density derivatives of the cutoff around the Mott density $\rho_{c}^{\mathrm{Mott}}$, whose temperature dependence was parametrized according to the parabolic expansion
\begin{equation}
\rho_{c}^{\mathrm{Mott}} = a_{c} + b_{c} T + c_{c} T^2 
\label{eq:parabolic}
\end{equation}
proposed in Ref.~\cite{ropkeNPA2011}.

Consistently with the choices adopted in
Refs.~\cite{WangArXiV2025,WangPRC2024}, we impose chemical equilibrium as a
reference condition for the initial state of the system,
\begin{equation}
\mu_{c} = A_{c} \mu_{\rm nuc},
\label{eq:chemeq}
\end{equation}
where $\mu_{\rm nuc}$ is the nucleon chemical potential, being the same for neutrons and protons. Since the infrared cutoff is assumed to only depends on the total baryon density
$\rho_{b}$ (so that
$\tilde{\mu}_{c}=A_{c} \tilde{\mu}_{\rm nuc}$),
Eq.~\eqref{eq:chemeq} can be rewritten as
\begin{equation}
\mu_{c}^{\ast}
=
A_{c}\mu_{\rm nuc}^{\ast} + B_{c}
+ A_{c}U_{\rm nuc}\left(1-\eta\right),
\label{eq:chemeq_eta}
\end{equation}
where 
$\eta$ is the screening factor 
introduced in Eq.~\eqref{eq:eta},
accounting for the fact that the mean-field potential $U_{c} / A_{c}= \eta U_{\rm nuc}$ felt by each nucleon bound in clusters may be smaller than that experienced by free nucleons ($U_{\rm nuc}$) in the surrounding nuclear medium.  
Equation~\eqref{eq:chemeq_eta} uniquely determines the cluster mass fraction
\begin{equation}
X_{c}=A_{c} \dfrac{\rho_{c}}{\rho_{b}}    
\end{equation}
for each value of the total baryon density.  Chemical equilibrium, however, is not essential to the present approach and could be relaxed by allowing for arbitrary cluster chemical potentials. Here it serves only as a convenient reference for calibrating the in-medium effects against benchmark (semi-)microscopic calculations. 

Within these assumptions and under different scenarios, in the following sections, we determine  both the spinodal boundary and the direction of the instability in the space of density fluctuations. 

As a first application, in Sec.~\ref{subsec:deuterons} we consider a simplified scenario in which only a single light cluster species is included as an explicit degree of freedom. This choice provides a validation of the present calculations against 
the formalism discussed in Sec.~\ref{subsec:hydro_vlasov} and serves as a benchmark for extending the analysis in Sec.~\ref{subsec:deuteron_alpha} to the simultaneous inclusion of multiple
cluster species.

\subsection{Nuclear matter with a single cluster species}
\label{subsec:deuterons}

In this section, following Ref.~\cite{WangPRC2024}, we include only deuterons as
explicit cluster degrees of freedom. Although at low temperatures both
microscopic quantum statistical (QS) and relativistic mean-field (RMF)
calculations~\cite{typelPRC2010} predict a clear dominance of $\alpha$ particles, for SNM at
temperatures $T \gtrsim 5~\mathrm{MeV}$—the regime of interest here—the
thermodynamics is largely governed by two-body correlations~\cite{WuJLTP2017}.
{However, the impact of the presence of $\alpha$ particles on the instability features will be addressed in Sec.~\ref{subsec:deuteron_alpha}.}

Under the assumption of a single cluster species, in the following subsections we perform a sensitivity study of the functional form adopted for the infrared cutoff [Eq.~\eqref{eq:EMt}] (Sec.~\ref{sec:results}\ref{subsec:sensitivity_cutoff}) and provide a direct comparison with the linear-response analysis based on the Vlasov equations of Ref.~\cite{WangPRC2024} (Sec.~\ref{sec:results}\ref{subsec:comparison}).

\begin{figure}[H]
\centering
\includegraphics[width=.42\textwidth]{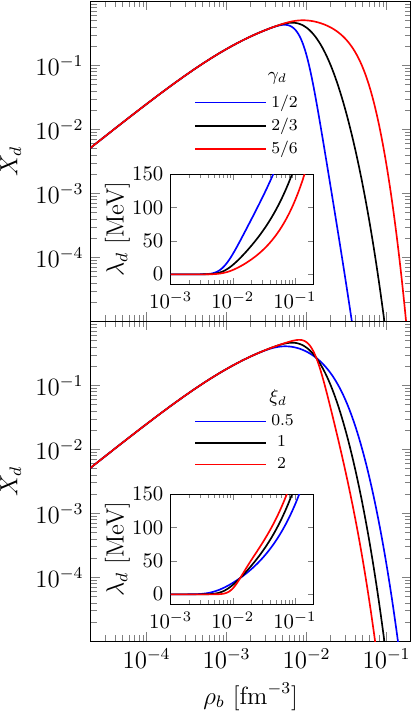}
\caption{Deuteron mass fraction $X_{d}$ as a function of the total baryon density
$\rho_{b}$ for different values of $\gamma_{d}$ (top panel) and $\xi_{d}$
(bottom panel). The insets show the density dependence of the corresponding
infrared cutoff parametrizations.}
\label{fig:xd_gamma_xi}
\end{figure}

\singpart{Sensitivity study to the cutoff parameterization}
\label{subsec:sensitivity_cutoff}
We first examine the sensitivity of the thermodynamic stability to two parameters of the cutoff parametrization: the stiffness at high densities, controlled by $\gamma_{d}$, and the smoothness of the transition around the
Mott density, governed by $\xi_{d}$. For simplicity, we consider the standard case, $\eta=1$, for the mean-field screening factor.  From Eq.~\eqref{eq:chemeq_eta} the deuteron mass fraction is obtained, whose resulting baryon density dependence is shown in
Fig.~\ref{fig:xd_gamma_xi}. In addition to the reference choice adopted in
Ref.~\cite{WangPRC2024} ($\gamma_{d}=2/3$, $\xi_{d}=1$), we consider two
alternative values for each parameter, while keeping $\nu_{d}=2$ and the Mott density $\rho_{d}^{\mathrm{Mott}}$ fixed to the values used in Ref.~\cite{WangPRC2024}. 
All
curves are obtained at a fixed temperature, $T=8~\mathrm{MeV}$.


As a general feature, all curves exhibit the characteristic rise-and-fall behavior of the deuteron mass fraction, with a maximum reached around the Mott density. Below this density, where the kinetic-energy cutoff approaches zero, all lines tend to coincide. Beyond the Mott density, $X_{d}$
decreases as a consequence of the (monotonically) increasing behavior of the cutoff
$\lambda_{d}$ with baryon density. In both panels of
Fig.~\ref{fig:xd_gamma_xi}, the rate at which the deuteron fraction is
suppressed above the Mott density directly reflects the behavior of the
cutoff  (see insets). In the upper panel, a smaller (larger) value of $\gamma_{d}$
corresponds to a stronger (weaker) density dependence of the in-medium
repulsion induced by $\lambda_{d}$ and therefore to a stiffer (softer)
suppression of $X_{d}$ at high densities. At the same time, the results are
sensitive to the smoothness of the cutoff around the Mott density, as
illustrated in the lower panel. A sharper transition leads to a more abrupt
dissolution of deuterons beyond the Mott density, 
despite the larger maximum
value of $X_{d}$ reached at $\rho_{d}^{\mathrm{Mott}}$.

\begin{figure}[H]
\centering
\includegraphics[width=.4\textwidth]{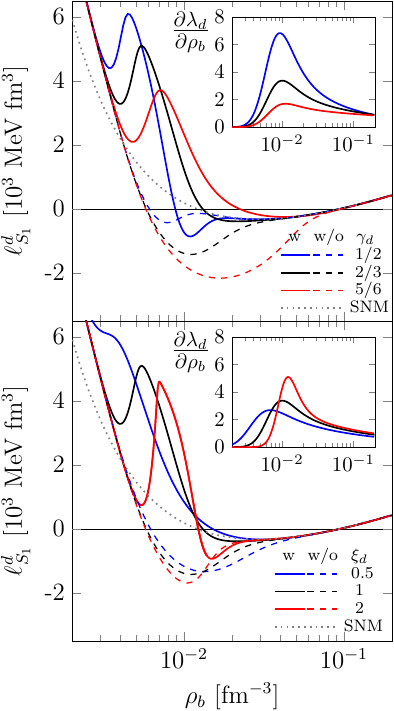}
\caption{Eigenvalue $\ell_{S_1}^{d}$ (see text) as a function of the total baryon density $\rho_{b}$ for different values of $\gamma_{d}$ (top panel) and $\xi_{d}$ (bottom panel). The full calculations (solid lines) are compared with the case in which the density dependence of the cutoff parameterizations is neglected 
in the stability analysis ($\Phi^{dj} = 0$) 
(dashed lines) and with the pure nucleonic case (SNM, grey dotted line). The insets show the corresponding behavior of the density derivative of the infrared cutoff parametrizations, 
expressed in the same units as in the main panels.
} 
\label{fig:eigenvalue_gamma_xi}
\end{figure}

To determine the spinodal boundary, we diagonalize the free-energy curvature matrix
[Eq.~\eqref{eq:c2lambda}], yielding two isoscalar
($\delta\rho_{n}=\delta\rho_{p}$) eigenmodes with eigenvalues
$(\ell_{S_1}^{d}, \ell_{S_2}^{d})$,\begin{equation}
\ell_{S_{1,2}}^{d}
=
\frac{\mathbb{C}_{11}+\mathbb{C}_{22}
\mp \sqrt{\left(\mathbb{C}_{11}-\mathbb{C}_{22}\right)^{2}
+4\mathbb{C}_{12}^{2}}}{2},
\label{eq:eigenvalue}
\end{equation}
and the corresponding eigenvectors in the
$(\delta\rho_{S},\delta\rho_{d})$ subspace.  The vanishing of the lower eigenvalue $\ell_{S_1}^{d}$ 
(corresponding to the minus sign in Eq.~\eqref{eq:eigenvalue}) determines the spinodal boundary,
while the associated eigenvector specifies the direction of the
instability. The ratio of its 
components 
is obtained as 
\begin{equation}
\left( \frac{\delta \rho_{S}}{\delta \rho_{d}} \right)_{1}
=
\frac{\ell_{S_{1}}^{d}-\mathbb{C}_{22}}{\mathbb{C}_{12}}
\label{eq:eigenvector}
\end{equation}
and is particularly instructive: positive (negative) values indicate that nucleon and deuteron density fluctuations evolve in phase (out of phase), respectively. 
This feature indicates that light clusters can either cooperate with nucleons in developing the mean-field instability or be pushed towards the lower density regions, where in-medium effects have less impact, thus
reinforcing a distillation-like mechanism~\cite{baranPREP2005,burrelloPRC2014}.

In Fig.~\ref{fig:eigenvalue_gamma_xi}, for the same  values of $\gamma_{d}$ (top
panel) and $\xi_{d}$ (bottom panel) as in Fig.~\ref{fig:xd_gamma_xi}, and at the temperature $T = 8$ MeV, we show the density dependence of the eigenvalue, $\ell_{S_1}^{d}$, 
whose negative values signal 
mechanical instability. To assess the role of in-medium effects, the full results (solid lines) are compared with
illustrative calculations (dashed lines) obtained by neglecting the density dependence of the cutoff,
($\Phi^{dj} = 0$). 
The corresponding eigenvalue in pure nucleonic matter (labeled as SNM) is also shown as a gray dotted line. The latter exhibits the standard 
unstable density
region driven by the mean 
field potential, bounded by the vanishing of the eigenvalue, beyond which mechanical stability is restored. 

When clusters are included, a
richer scenario emerges. At low densities, the curves with clusters are shifted above the SNM result due to the small cluster fraction (level density), which leads to large derivatives of the (effective) chemical potentials. With increasing density, the cluster mass fraction grows and the magnitude of the eigenvalues is reduced by the attractive mean-field contribution, which is stronger for clusters than for nucleons. As a result, when in-medium effects
are neglected (dashed curves), the instability region is significantly
enhanced and extends over a wider density range than in pure nucleonic matter. At higher densities, clusters are progressively suppressed and eventually dissolve into the medium, causing all curves to converge.

When in-medium effects are fully included, the solid curves 
begin to deviate from the dashed ones around the Mott density, 
where a characteristic bump appears. This feature originates from the repulsion induced by Pauli-blocking effects and is directly linked to the density derivative of the cutoff (see inset). 
This contribution adds a positive (kinetic) contribution to 
the chemical potential derivatives, partially
counterbalancing the mean-field attraction and thus generally reducing the instability.
As a result of this particular behavior, 
when in-medium effects are included, the extension of the unstable region becomes closer to that of SNM. 
In this case, the system may even exhibit two distinct instability regions, separated by an intermediate mechanically metastable domain, 
as discussed in
Refs.~\cite{RopkeNPA2018,VoskresenskyPPNP2023,WangPRC2024}.

Nevertheless, the general scenario depends sensitively on the cutoff
parametrization: 
the smoothness parameter primarily controls the magnitude of the
bump, while having only a minor impact on the onset of the instability
region. By contrast, the density-slope parameter $\gamma_{d}$ plays a crucial
role in determining its extent. 

\begin{figure}[tbp!]
\centering
\includegraphics[width=.4\textwidth]{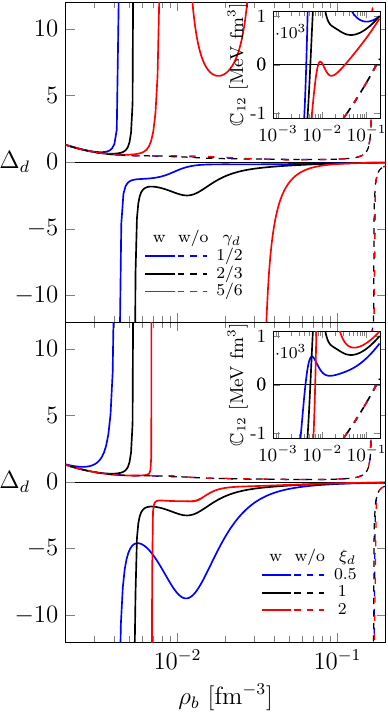}
\caption{The quantity $\Delta_{d}$ (see text) as a function of the total baryon
density $\rho_b$ for different values of $\gamma_{d}$ (top panel) and $\xi_{d}$ (bottom panel). The full calculations (solid lines) are compared with the case in which the density dependence of the cutoff parameterizations is neglected 
in the stability analysis ($\Phi^{dj} = 0$) 
(dashed lines). The insets show the corresponding behavior of the isoscalar-cluster component of the free-energy curvature matrix $\mathbb{C}_{12}$.}
\label{fig:eigenvector_gamma_xi}
\end{figure}
The direction of the instabilities and the nature of the associated modes can be further characterized by analyzing the eigenvectors of the free-energy curvature
matrix. 
Following  Ref.~\cite{WangPRC2024}, 
we introduce the normalized quantity
\begin{equation}
\Delta_{d}
=
\left ( \frac{\delta\rho_{S}}{\delta\rho_{d}} \right)_{1} \,
\frac{\rho_{d}}{\rho_{S}},
\label{eq:deltad}
\end{equation}
which is displayed in Fig.~\ref{fig:eigenvector_gamma_xi} as a function of the
total baryon density $\rho_{b}$, for both calculation schemes discussed above,
namely neglecting or fully including the density dependence of the cutoff, for the temperature $T = 8$ MeV.

At densities below the Mott density, deuteron fluctuations are found to be in phase with nucleonic ones. When the density dependence of the cutoff is neglected, this behavior persists up to densities above
$10^{-1}\,\mathrm{fm}^{-3}$, 
i.e., beyond the spinodal region. It should be noted that, according to Eq.~\eqref{eq:eigenvector}, the $\Delta_d$ ratio changes sign,
exhibiting a divergent behavior, 
when the $\mathbb{C}_{12}$ component of the free-energy curvature matrix crosses zero (see the inset). 

When the density-dependent cutoff is considered, the $\Delta_d$ ratio changes significantly above the Mott density. 
The divergence shifts to lower densities and deuteron fluctuations become out of phase with respect to nucleonic ones already within the spinodal region. 
While the density at which this feature occurs depends only weakly on the smoothness parameter $\xi_{d}$ of the cutoff, it exhibits a strong
dependence on the stiffness parameter $\gamma_{d}$ 
(see in particular the soft case, corresponding to $\gamma_d = 5/6$, in the upper panel, where the out-of-phase behavior only appears in a limited density range inside the spinodal region). 


\singpart{Thermodynamical stability and linearized Vlasov approach}
\label{subsec:comparison}
In this section, we compare the thermodynamic stability analysis presented above with the linear-response study based on the collisionless Vlasov
approach of Ref.~\cite{WangPRC2024}, adopting the same reference choice
($\gamma_{d}=2/3$, $\xi_{d}=1$) for the cutoff parametrization.
Our focus here, however, is not on the dynamical growth rates of density
fluctuations, but rather on the extent of the instability region in the
phase diagram. This region is delimited by the spinodal boundary, which in the
thermodynamic framework is determined by the vanishing of the free-energy
curvature, i.e., by the zeros of the isoscalar eigenvalue $\ell_{S_{1}}^{d}$.
In the Vlasov case, the spinodal border corresponds to the density values where the determinant of the matrix in
Eq.~\eqref{eq:a2lambda_tilde} vanishes. 
\begin{figure}[H]
\centering
\includegraphics[width=.42\textwidth]{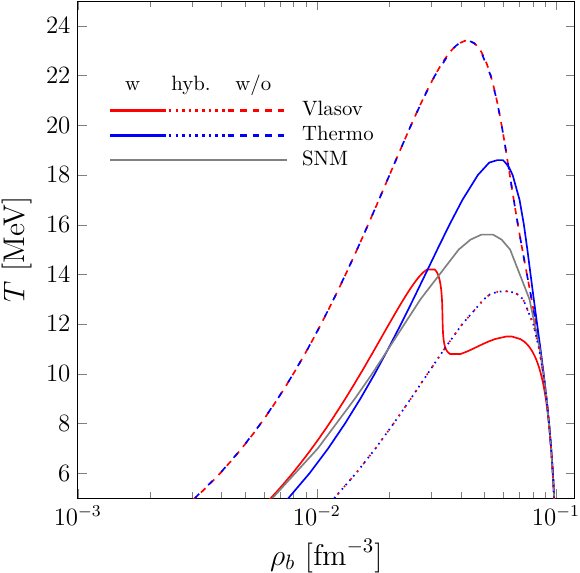}
\caption{Spinodal border in the $(\rho_{b},T)$ plane for nuclear matter with deuterons under different prescriptions:
(i) 
neglecting in-medium effects ($\Phi^{dj} = 0$) 
(dashed lines);
(ii) ``hybrid'' case (dash-dotted lines);
and (iii) full inclusion of in-medium effects (solid lines).
The red curves correspond to the linear-response (Vlasov) results of
Ref.~\cite{WangPRC2024}, while the blue curves represent the present
thermodynamic stability analysis. The spinodal boundary for pure nucleonic matter (SNM, gray) is shown for reference.}
\label{fig:spinodal_Vlasov_thermo}
\end{figure}

Figure~\ref{fig:spinodal_Vlasov_thermo} shows the spinodal border in the $(\rho_{b},T)$ plane for nuclear matter with deuterons under different prescriptions:
(i) 
neglecting in-medium effects ($\Phi^{dj} = 0$) 
(dashed lines);
(ii) ``hybrid'' case (dotted lines);
and (iii) full inclusion of in-medium effects (solid lines).
The red curves correspond to the linear-response (Vlasov) results plotted in Fig.~1 of
Ref.~\cite{WangPRC2024}, while the blue curves represent the
thermodynamic stability analysis. The spinodal boundary for pure nucleonic matter (SNM, gray) is shown for reference.
It is worth noting that the Coulomb interaction is expected to shrink the spinodal region relative to the uncharged case considered here, since it penalizes proton-density fluctuations and favors finite-size structures over bulk liquid–gas separation~\cite{DucoinNPA2007}. A quantitative assessment of this effect, including the contribution of protons bound in light clusters, would, however, require an explicit extension of the present framework and is therefore beyond the scope of this work.

As already highlighted in Ref.~\cite{WangPRC2024}, 
the explicit inclusion of light clusters is found to substantially modify the extent of the spinodal region. Indeed, when in-medium effects are neglected (dashed lines), the instability domain widens significantly. This behavior reflects the enhanced attractive contribution arising from the deuteron mean-field potential, which enters the
$F_{0}^{dd}$ term. In contrast, in the ``hybrid'' case (dotted lines), a pronounced reduction of the unstable region is obtained, since Pauli-blocking effects increase the deuteron kinetic energy and counteract the mean-field attraction. Remarkably,
once in-medium effects are fully included through the $\tilde{F}^{jl}$ (and $\tilde{\Phi}^{jl}$)
terms (red or blue solid curves), the resulting spinodal boundary of the composite system lies closer to that of pure nucleonic matter (gray solid curve). 
\begin{figure*}
\centering
\includegraphics[width=.8\textwidth]{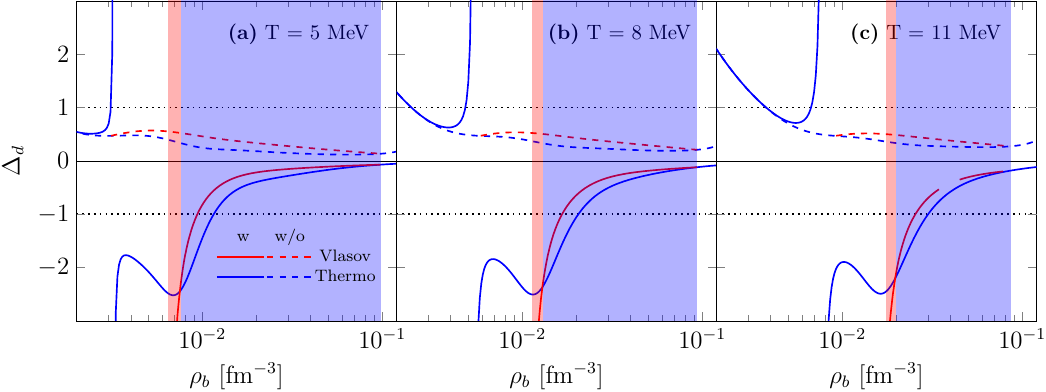}
\caption{The quantity $\Delta_{d}$ (see text) as a function of the total baryon density
$\rho_{b}$ for nuclear matter with deuterons, obtained by neglecting (dashed
lines) or fully including (solid lines) in-medium effects, at three different
temperatures. The red curves, drawn only inside the spinodal region, correspond
to the linear-response (Vlasov) results of Ref.~\cite{WangPRC2024}, while the blue curves represent the thermodynamic stability analysis. The shaded areas indicate the extent of the corresponding spinodal regions in the two approaches when in-medium effects are fully included. {The dotted lines, 
shown for reference, correspond to $|\Delta_{d}| = 1$.}}
\label{fig:eigvec_Vlasov_thermo}
\end{figure*}
The thermodynamic and Vlasov spinodals coincide both when the cutoff is treated as density independent (dashed curves) and in the hybrid case, $\tilde{F}^{jl} = \tilde{\Phi}^{jl} = 0$ (dotted curves), because no rearrangement contributions arise and the long-wavelength limit of the Vlasov dispersion relation reduces to the standard Landau stability condition based on the free-energy curvature. This was already proved in Sec.~\ref{subsec:hydro_vlasov} by the fact that, in these two cases,  the zeros of the determinant of $\mathbb{C}$ 
and $\mathbb{A}$
manifestly emerge at the same density values. 

The discrepancy observed in the full calculations (solid curves) arises solely 
from the additional terms, with different density behavior, in the chemical potentials ($\tilde{\mu}_{j}$) and in the single-particle energies ($\tilde{\varepsilon}_{j}$). One observes that the density dependence of the cutoff quenches the collisionless
growth of fluctuations in the Vlasov framework, lowering the critical temperature, while it allows thermodynamic instability to persist up to higher temperatures. This reflects the fact that the thermodynamic and Vlasov criteria probe, in general, different notions of instability: the former identifies the loss
of convexity of the free energy, whereas the latter governs the dynamical growth
of density fluctuations, 
driven by the single-particle potential. The density dependence of the cutoff generates an additional contribution to the pressure, $\tilde{P}$, that is responsible for the difference observed between ``dynamical'' and ``thermodynamical'' instabilities. 
Consequently, when the cutoff is density dependent, a
region may emerge in which phase separation is thermodynamically
favored but the dynamical
amplification of density fluctuations is inhibited.
However, we stress that this behavior mainly concerns the high temperature region ($T \gtrsim  11$ MeV), whereas the two approaches yield quite similar results elsewhere, pointing to 
modest effects of $\tilde{P}$. 
{This agreement further suggests that the instability features discussed in Ref.~\cite{WangPRC2024}
reflect intrinsic thermodynamic properties of the medium.}

In Fig.~\ref{fig:eigvec_Vlasov_thermo}, we further compare the quantity
$\Delta_{d}$ defined in Eq.~\eqref{eq:deltad}, as obtained in the present
thermodynamic analysis, with the corresponding Vlasov results (drawn only inside the spinodal region) shown in Fig.~4
of Ref.~\cite{WangPRC2024}. The extents of the spinodal regions predicted by the
two approaches in the case of full inclusion of in-medium effects are
highlighted by the shaded areas. It should be noted that, although close, the dashed curves do not exactly coincide. Indeed, the Vlasov results are obtained from the full solution of the isoscalar--cluster block of
Eq.~\eqref{eq:vlasov_linear}, which explicitly includes the Lindhard functions $\chi_{q}$ and $\chi_{d}$. Only in the limiting case $\chi_{q}=\chi_{d}=1$, namely at the border of the dashed spinodal regions of Fig.~\ref{fig:spinodal_Vlasov_thermo},
the dashed lines  coincide
in Fig.~\ref{fig:eigvec_Vlasov_thermo}.
However, 
also in the case 
in which 
the density-dependent cutoff 
is fully taken into account, both in the stability analysis and in the dynamical response, 
despite the differences introduced by the additional terms, 
the two approaches show good overall agreement at the temperatures considered here. In particular, they predict not only the same sign—thus the same in-phase or out-of-phase character of the density fluctuations—but also comparable magnitudes of $\Delta_{d}$ across the
instability region.
From Fig.~\ref{fig:eigvec_Vlasov_thermo} it can also be observed that the relative weight of the cluster oscillations presents an initial increasing trend as a function of the density, in connection with the cluster abundance; indeed the quantity $\Delta_d$ decreases, especially at higher temperature, reaching values lower than 1, pointing to an increasing relative contribution of cluster oscillations with respect to nucleons. However, close to the Mott density, the trend is reversed: deuteron oscillations vanish (causing the divergence observed in the figure) and then change sign. However, inside the spinodal region, the contribution of cluster oscillations is significant again, though they move out-of-phase with respect to nucleons, signaling the occurrence of sizable migration effects. 

\subsection{Nuclear matter with deuterons and alphas}
\label{subsec:deuteron_alpha}
We finally examine the composition and thermodynamic stability of nuclear matter including both deuterons and $\alpha$ particles. 
To this end, in Sec.~\ref{sec:results}\ref{subsec:d_a_parameterizations} we first specify the parametrizations adopted for the Mott densities and the associated infrared cutoffs of both cluster species. 
This allows us to delineate the general features of the system across the full phase diagram (Sec.~\ref{sec:results}\ref{subsec:spinodal_d_a}). 
Finally, in Sec.~\ref{sec:results}\ref{subsec:sensitivity_MF}, we investigate the sensitivity of the results to different assumptions on the binding energies $B_{c}$ ($c=d,\alpha$) and on the mean-field screening factor $\eta$.

\singpart{Cutoff parameterizations and characterization of the instability} \label{subsec:d_a_parameterizations}
A fully microscopic determination of the Mott momenta as functions of density and temperature would
require solving the in-medium many-body Schr\"odinger
equation~\cite{ropkeNPA2011}, which becomes computationally demanding for clusters heavier than deuterons. We therefore fix these quantities by calibrating the in-medium effects against benchmark calculations according to
two alternative strategies: (i) a phenomenological parametrization adjusted to reproduce the cluster mass fractions of RMF calculations of
Ref.~\cite{typelPRC2010}, labeled as \textit{stiff} in the following; and (ii) the semi-microscopic phase-space
excluded-volume prescription of Ref.~\cite{WangArXiV2025} {(using the set (i) of threshold parameters $\mathbf{F}^{\rm cut}$),} 
hereafter denoted as \textit{soft}. 
The latter provides an effective estimate of 
Mott density and (density dependent) Mott momentum as a function of the temperature of the surrounding nuclear medium, 
thus capturing the essential features of the more sophisticated in-medium Schr\"odinger equation, at a significantly reduced computational cost.

A clarification is in order in this respect. The cluster mass fractions obtained within RMF (and QS) calculations include not only bound states but
also contributions from continuum correlations. These correlated scattering states enter the thermodynamics through virial-like or many-body correlation
terms and thus affect the total baryonic composition beyond  {the present description},
which focuses on the survival and dissolution only of bound clusters. 
Furthermore, the RMF benchmark calculations also include tritons and
${}^{3}$He. 
In the stiff case, calibrated to reproduce the RMF deuteron and
$\alpha$ fractions, the density dependence of the cutoff is {consequently}
particularly \textit{stiff}, as it effectively incorporates additional
physical contributions such as continuum correlations and the
presence of other cluster species. By contrast, within the
phase-space excluded-volume approach,
 the omission of tritons and ${}^{3}$He may
further 
enhance the deuteron and $\alpha$ abundances obtained with
the soft parametrization. The differences between the soft and stiff
cutoffs, especially pronounced in the case of $\alpha$ particles, can therefore be traced back to the absence, in the former, of
continuum correlations and additional light clusters that are effectively
embedded in the latter.


The values of the parameters $\beta_{c}$, $\gamma_{c}$, $\xi_{c}$ and $\nu_{c}$ obtained by fitting the functional forms of Eqs.~\eqref{eq:EMt} and~\eqref{eq:smoothing}, for the temperature $T = 8$ MeV, according to the prescriptions
described above, are listed in Table~\ref{tab:cut_off}, together with the choice adopted in Ref.~\cite{WangPRC2024}. In Table~\ref{tab:cut_off}, we also report the coefficients $a_{c}$, $b_{c}$, and $c_{c}$ of Eq.~\eqref{eq:parabolic} obtained to reproduce the deuteron and $\alpha$ Mott densities (or the maxima of the deuteron and $\alpha$ mass fractions) at $T = 5$, $8$, and $11$ MeV, for the soft (or stiff) parametrizations. 
\begin{table*}[tbp!]
\centering
\begin{tabular}{cc @{\hskip 1.2cm} c @{\hskip 1.2cm} c c
@{\hskip 1.2cm} c c}
\toprule
Parameter & Units & Ref.~\cite{WangPRC2024} & \multicolumn{2}{c}{\hspace{-1.2cm}soft}& \multicolumn{2}{c}{stiff} \\ & & d & d & $\alpha$  & d & $\alpha$\\
\midrule
$\beta_{c}$ & $10^{2}$ fm$^{3\gamma_{c}}$ MeV & $4.400$ & $1.5271$ & $2.8919$ & $9.5452$ & $14.055$ \\
$\gamma_{c}$ & -- & $2/3$ & $0.5337$ & $0.6296$ & $1.0091$ & $0.88746$ \\
$\xi_{c}$ & -- & $1$ & $3.0954$ & $1.8788$ & $24.377$ & $66.624$ \\
$\nu_{c}$ & -- & $2$ & $1.5589$ & $1.5369$ & $0.00385$ & $0.00310$ \\
\midrule
$a_{c}$ & $10^{-3}$ fm$^{-3}$ & $0.5765$ & $0.5156$ & $1.7278$ & $10.394$ & $8.1979$ \\
$b_{c}$ & $10^{-4}$ fm$^{-3}$ MeV$^{-1}$ & $6.5443$ & $1.6778$ & $11.806$ & $-7.6642$ & $15.528$ \\
$c_{c}$ & $10^{-5}$ fm$^{-3}$ MeV$^{-2}$ & $1.2491$ & $2.2222$ & $2.2778$ & $85.786$ & $33.269$ \\
\bottomrule
\end{tabular}
\caption{Parameters obtained by fitting the deuteron and alphas Mott energies (maxima of mass fractions) from Ref.~\cite{WangArXiV2025} (Ref.~\cite{typelPRC2010}), according to the functional form of Eqs.~\eqref{eq:EMt} and~\eqref{eq:smoothing}, for the soft (stiff) parameterization. For comparison, the parametrization of Ref.~\cite{WangPRC2024}, available only for deuterons, is also listed.}  
\label{tab:cut_off}
\end{table*}

\begin{figure}[tbp!]
\centering
\includegraphics[width=.3\textwidth]{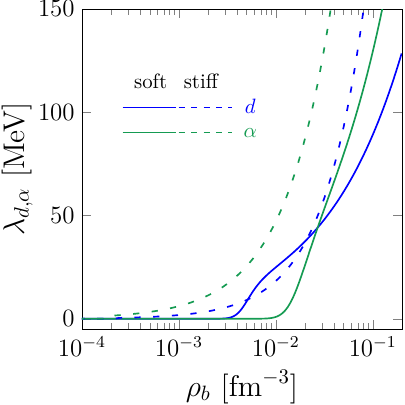}
\caption{Kinetic energy cutoff for deuterons (blue) and $\alpha$ particles (green) as function of the total baryonic density $\rho_{b}$ as obtained at $T = 8$ MeV, for the soft (solid) or stiff (dashed) parameterizations (see Table~\ref{tab:cut_off}). 
}
\label{fig:cutoff_stiff_soft}
\end{figure}

The corresponding curves are shown in
Fig.~\ref{fig:cutoff_stiff_soft}: blue (green) lines denote deuterons ($\alpha$
particles), and dashed (solid) lines indicate stiff (soft)
parametrizations.
We note that the density behavior of the cutoff cannot be inferred from the value of the slope parameter $\gamma_{c}$ alone. While the sensitivity analysis varied $\gamma_{d}$ with the other parameters
fixed, thereby isolating its specific effect on the density dependence
of $\lambda_{d}$, the soft and stiff parametrizations of
Table~\ref{tab:cut_off} arise from correlated global fits in which
$\beta_{c}$, $\gamma_{c}$, $\xi_{c}$, and $\nu_{c}$ are simultaneously
adjusted. The magnitude and local slope of the cutoff are therefore governed by
the combined action of all parameters, which explains the apparent
inversion of the $\gamma_{d}$ trends in
Fig.~\ref{fig:cutoff_stiff_soft} 
(or Table~\ref{tab:cut_off}) with respect to the analysis
shown in the inset of Fig.~\ref{fig:xd_gamma_xi}. Moreover, it is worth clarifying why the smoothing parameters entering the stiff parametrization lie well beyond the range explored in the sensitivity analysis of Section~\ref{sec:results}\ref{subsec:sensitivity_cutoff}. In this case, the cutoff 
is reconstructed from benchmark cluster mass fractions, which implicitly embed additional physical contributions, such as continuum correlations. As a consequence, the effective Mott density is shifted toward lower values, and the smoothing parameters effectively absorb physical effects beyond those considered in the adopted cutoff parametrization.

\begin{figure}[tbp!]
\centering
\includegraphics[width=.4\textwidth]{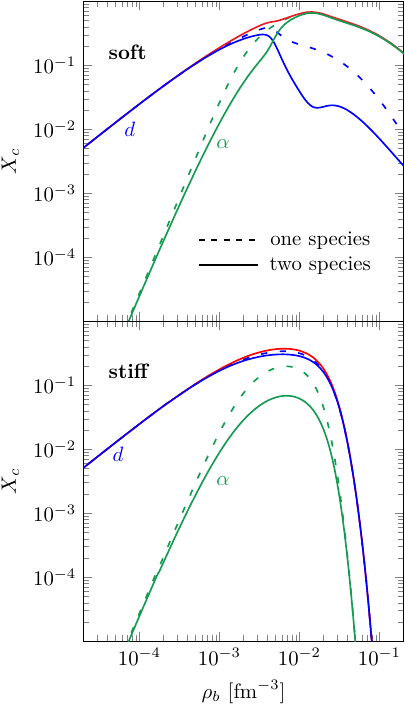}
\caption{Cluster mass fractions $X_{c}$ ($c=d,\alpha$) as functions of the total baryon density $\rho_{b}$ at $T = 8$ MeV. Dashed lines correspond to
calculations in which only one cluster species is included at a time,
whereas solid lines represent the simultaneous inclusion of both species.
For the latter case, the total cluster fraction $X_{d+\alpha} = X_{d} + X_{\alpha}$ is also shown (red lines).
Two cutoff parametrizations are considered: soft (top panel) and stiff
(bottom panel).} 
\label{fig:X_c_soft_vs_stiff_d_and_a}
\end{figure}

Figure~\ref{fig:X_c_soft_vs_stiff_d_and_a} shows the corresponding cluster mass fractions $X_{c}$ ($c=d,\alpha$) as functions of the
total baryon density $\rho_{b}$, deduced according to Eq.~\eqref{eq:chemeq_eta}, at $T = 8$ MeV. Dashed lines correspond to calculations in
which only a single light cluster species is included, whereas solid lines represent the case where both species are simultaneously considered. The two
cutoff parametrizations are explored: soft (top panel) and stiff (bottom panel).

First, already in the single-species case, the two cutoff parametrizations predict quite different cluster mass fractions beyond
the respective Mott densities. As a general feature, apart from a slightly sharper suppression of the deuteron fraction immediately above its Mott
density, the soft parametrization leads to a smoother dissolution of
clusters. This behavior is consistent with the lower kinetic-energy cutoff
predicted by the soft parametrization (see
Fig.~\ref{fig:cutoff_stiff_soft}), particularly for $\alpha$ particles. As a consequence, in the stiff case the deuteron fraction (blue lines) remains systematically larger than that of $\alpha$ particles (green
lines). In contrast, in the soft parametrization $X_{\alpha}$ exceeds
$X_{d}$ at sufficiently large densities, beyond an intermediate region where the two species may reach comparable abundances. When both species are included simultaneously, each individual cluster
fraction is reduced due to the mutual competition for (low-momentum) quasi-particle states in the medium. This reduction is in general more pronounced for the subdominant species. Nevertheless, the total cluster fraction
$X_{d + \alpha} = X_{d} + X_{\alpha}$ (red lines) always exceeds the
corresponding single-species result and closely follows the dominant component. 

\begin{figure}[tbp!]
\centering
\includegraphics[width=.4\textwidth]{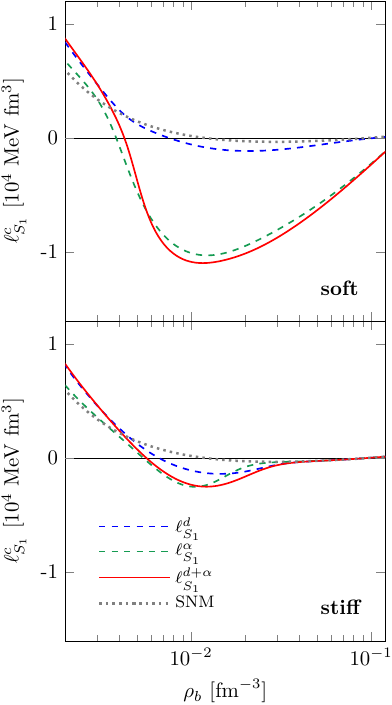}
\caption{Eigenvalues $\ell_{S_1}^{c}$ (see text) as function of the total baryon density $\rho_{b}$, at $T = 8$ MeV, 
in the case where the density dependence of the cutoff is neglected in the stability analysis ($\Phi^{cj}= 0$). 
Dashed blue (green) lines correspond to calculations in which only deuterons (alphas) are included at a time, whereas red solid lines represent the simultaneous inclusion of both species. The corresponding eigenvalue for pure nucleonic matter (gray dotted lines) is shown for reference. Two cutoff parametrizations are considered: soft (top panel) and stiff
(bottom panel). 
}
\label{fig:eigval_stiff_soft_NME}
\end{figure}

Figure~\ref{fig:X_c_soft_vs_stiff_d_and_a} is also helpful in interpreting
the results shown in Fig.~\ref{fig:eigval_stiff_soft_NME}, where—similarly
to Fig.~\ref{fig:eigenvalue_gamma_xi}—the density dependence of the
eigenvalue $\ell_{S_1}^{c}$, $c= d, \alpha, d+\alpha$ is displayed,  in the case where the density dependence of the cutoff is neglected 
in the stability analysis ($\Phi^{cj} = 0$).  
One observes that, consistently with the deuteron mass fractions shown in
Fig.~\ref{fig:X_c_soft_vs_stiff_d_and_a}, the eigenvalue
$\ell_{S_1}^{d}$ (dashed blue lines) takes similar values for the stiff and soft parametrizations. The pronounced difference observed in
$\ell_{S_1}^{d+\alpha}$ (red solid lines)—the lower isoscalar eigenvalue
in the two-species case and the only one that becomes negative,
thereby signaling the onset of mechanical instability—is therefore
mainly driven by the behavior of $\ell_{S_1}^{\alpha}$ (dashed green lines). The latter reflects the 
large differences in the $\alpha$ mass fractions
between the stiff and soft parametrizations. 
Moreover, $\ell_{S_1}^{d+\alpha}$ closely follows the dominant
single-species contribution, 
enhancing the instability in the intermediate-density region where both species reach sizable abundances. 

\begin{figure}[tbp!]
\centering
\includegraphics[width=.4\textwidth]{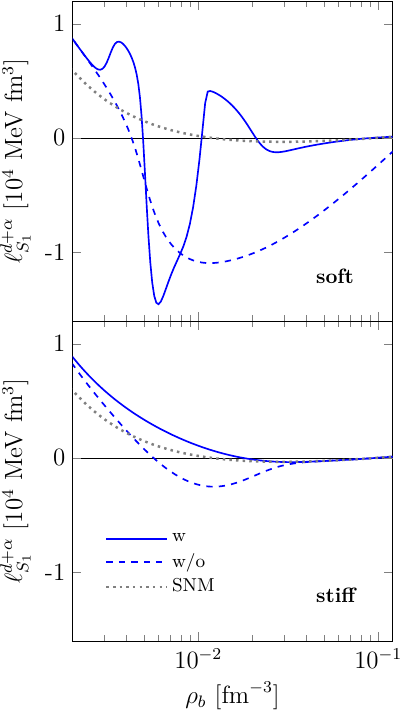}
\caption{Eigenvalue $\ell_{S_1}^{d+\alpha}$ (see text) as function of the total baryon density $\rho_{b}$, at $T = 8$ MeV, as obtained in the case of the simultaneous inclusion of both cluster species, for the two cutoff parametrizations
considered: soft (top panel) and stiff (bottom panel). Dashed (solid) lines correspond to calculations neglecting (fully including) in-medium effects. The corresponding eigenvalue for pure nucleonic matter (dotted lines) is shown for reference.} 
\label{fig:eval_soft_stiff}
\end{figure}
However, as already discussed in the sensitivity analysis related to
Fig.~\ref{fig:eigenvalue_gamma_xi}, in-medium effects substantially
modify both the onset and the magnitude of the instability. This is
illustrated in Fig.~\ref{fig:eval_soft_stiff}, where the 
eigenvalues shown in Fig.~\ref{fig:eigval_stiff_soft_NME} for the
two-species case  $\ell_{S_1}^{d+\alpha}$—neglecting the density dependence of the cutoff 
in the stability analysis ($\Phi^{cj} = 0$) 
(dashed lines)—are 
compared with the corresponding full
calculations including in-medium effects (solid lines), 
for both the soft (top panel) and stiff
(bottom panel) parametrizations.

In the stiff case, the usual suppression of the instability is recovered, approximately restoring both the magnitude and the extent of the standard pure nucleonic matter spinodal region. The situation is
markedly different in the soft case. 

There, 
the system first enters the wide (dashed) spinodal region with even larger magnitude, and subsequently reenters the standard density domain associated with the SNM instability, again with comparable magnitude and extent. 
This additional low-density instability branch can be traced back to the large $\alpha$-particle abundances predicted by the soft parametrization, which allow the cluster mean-field attraction to remain sufficiently strong despite the repulsive effect induced by the density-dependent cutoff around the Mott region. 

\begin{figure*}
\includegraphics[width=.453\textwidth]{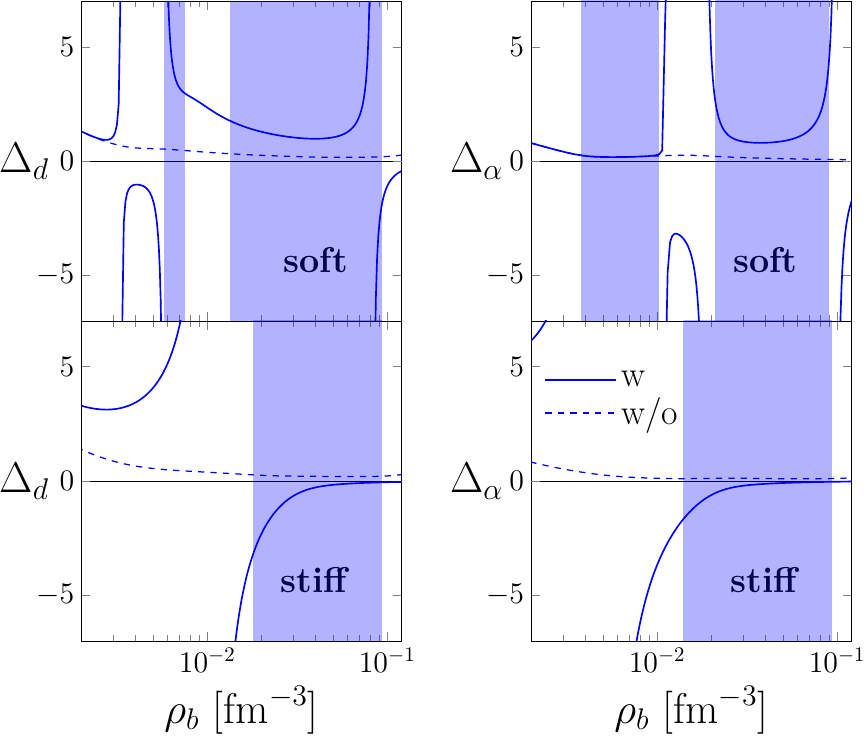} \qquad \qquad
\includegraphics[width=.433\textwidth]{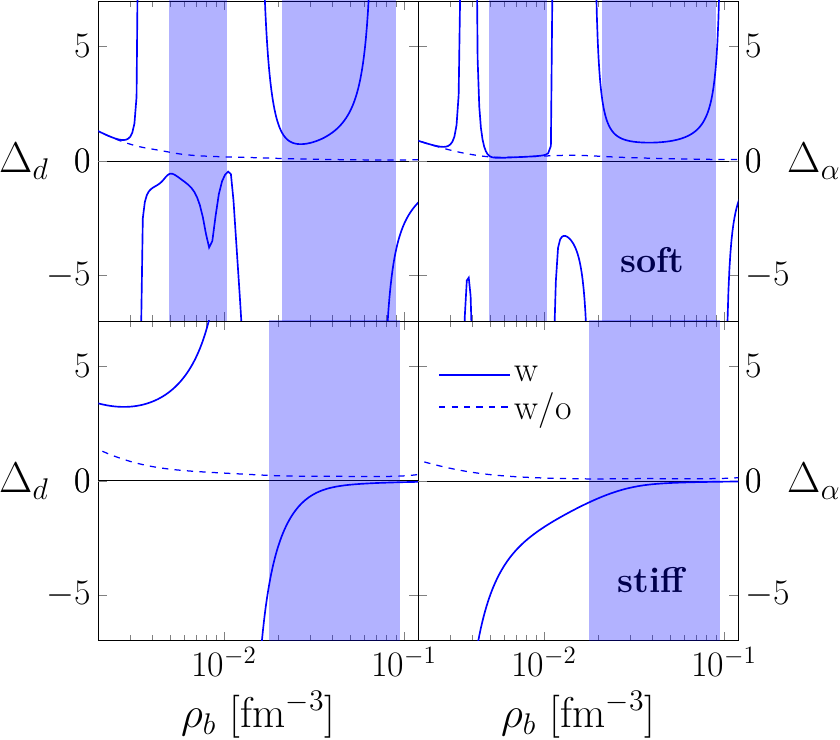}
\caption{Quantities $\Delta_{d}$ and $\Delta_{\alpha}$ (see text) as functions
of the total baryon density $\rho_{b}$ at $T=8$ MeV. Dashed (solid) lines
correspond to calculations neglecting (fully including) in-medium effects.
The left panels refer to calculations in which only one light cluster species
($d$ for $\Delta_{d}$, $\alpha$ for $\Delta_{\alpha}$) is included at a time, whereas the right panels show the case where both species are simultaneously considered. The top (bottom) panels correspond to the soft (stiff) cutoff
parametrization. Shaded areas indicate the extent of the spinodal region in
the case of full inclusion of in-medium effects.} 
\label{fig:evec_soft_vs_stiff_d_and_a}
\end{figure*}

To finally characterize the nature of the unstable modes in the space of density fluctuations, let us look at the quantities $\Delta_{c}$, defined as
\begin{equation}
\Delta_{c} = \left( \dfrac{\delta \rho_{S}}{\delta \rho_{c}} \right)_{1} \dfrac{\rho_{c}}{ \rho_{S}}, \qquad c = d, \alpha      
\end{equation}
in analogy to Eq.~\eqref{eq:deltad}. These quantities are shown in Fig.~\ref{fig:evec_soft_vs_stiff_d_and_a} as functions of the total baryon density $\rho_{b}$ at $T=8$ MeV. 
Dashed (solid) lines correspond to calculations neglecting (fully including) in-medium effects.
The left two-column panels refer to calculations in which only one light cluster species ($d$ for $\Delta_{d}$, $\alpha$ for $\Delta_{\alpha}$) is included at a time, whereas the right two-column panels show the case where both species are simultaneously
considered. The top (bottom) panels correspond to the soft (stiff) cutoff
parametrization. Shaded areas indicate the extent of the spinodal region in
the case of full inclusion of in-medium effects. 

Once again, significant differences emerge depending on the cutoff
parametrization adopted to model in-medium effects. In the stiff case,
for a single cluster species, the overall picture is similar to that
discussed in Fig.~\ref{fig:eigvec_Vlasov_thermo} and in the linear
response analysis of Ref.~\cite{WangPRC2024}: the steep density
dependence of the cutoff drives clusters (either deuterons or
$\alpha$ particles) to fluctuate out of phase with nucleons, pushing
them toward lower-density regions while nucleonic instabilities grow.
This contrasts with the in-phase behavior obtained when in-medium
effects are neglected. 
In the two-species case, the stiff scenario remains essentially unchanged,
owing to the dominance of deuterons across the entire density range.
The inclusion of $\alpha$ particles leaves the deuteron fluctuations
$\Delta_{d}$ in the right panel practically unaffected. Conversely, the
presence of deuterons modifies the magnitude of $\Delta_{\alpha}$,
shifts the associated divergence to lower densities, and slightly
reduces the extent of the instability region (shaded area), while
preserving the out-of-phase character of both cluster fluctuations
inside the spinodal region.

On the other hand, the soft parametrization considerably enriches the picture. 
First, already in the single-species case and in analogy
with the behavior observed for the red curve in the upper panel of
Fig.~\ref{fig:eigenvector_gamma_xi}, additional divergences appear
when in-medium effects are fully included. 

In nuclear matter containing only deuterons, 
in contrast to the stiff case, the out-of-phase behavior appears only immediately above the deuteron Mott density, which lies below the first entry into the spinodal region. At the onset of that region, deuterons again fluctuate in phase with nucleons. This behavior originates from the softer density dependence of the cutoff, which weakens Pauli-blocking repulsion and leads to larger deuteron fractions
(Fig.~\ref{fig:X_c_soft_vs_stiff_d_and_a}), thereby enhancing the
mean-field attraction and favoring in-phase motion.

For nuclear matter containing only $\alpha$ particles, the qualitative
picture is similar.
Nucleons and $\alpha$ particles fluctuate in
phase within the instability regions, as indicated by the positive
values of $\Delta_{\alpha}$ in the shaded areas.

The scenario becomes more intricate in the two-species case, where the
isoscalar-cluster sector involves three coupled densities
$(\delta\rho_{S},\delta\rho_{d},\delta\rho_{\alpha})$ and the normal modes
are therefore reoriented in density space.
The inclusion of deuterons leaves the qualitative behavior of
$\Delta_{\alpha}$ largely unchanged compared to the single-species
case, except near the deuteron Mott density, where the additional
repulsion induces a low-density out-of-phase behavior (outside the unstable region) and slightly
reduces the extent of the first instability region. Conversely, the
presence of $\alpha$ particles strongly affects $\Delta_{d}$. 

{As a result, when in-medium effects are fully included and both species are considered, in the soft case $\alpha$ particles drive deuterons out of phase in the first instability region, pushing them toward lower densities. On the other hand, the growth of instabilities in the standard spinodal region is primarily supported by their cooperation with nucleons.}

To conclude, if the density dependence of the cutoff is neglected 
in the stability analysis, 
the characterization of the unstable modes inside the spinodal region is straightforward: light clusters always move in phase with nucleons, thereby reinforcing the growth of density fluctuations and effectively acting as seeds for fragment formation. When in-medium effects are fully included, however, the picture becomes more complex. Pauli blocking introduces a repulsion that tends to prevent clusters from populating high-density regions. However, if this repulsion does not increase too steeply with density, as for the soft parameterization, a sufficiently large fraction of clusters may survive at higher densities, and their mean-field attraction can drive them to cooperate in the formation of intermediate-mass fragments.

\singpart{Thermal effects and spinodal region}
\label{subsec:spinodal_d_a}
The results discussed in Sec.~\ref{sec:results}\ref{subsec:d_a_parameterizations} focus on
$T = 8$ MeV. To explore the spinodal region over the full
$(\rho_{b},T)$ phase diagram, thermal effects must be incorporated by
accounting for the temperature dependence of the cutoff. In the present framework, this dependence is introduced through the Mott densities,
parametrized via the parabolic expansion of
Eq.~\eqref{eq:parabolic}, 
whose parameters $a_{c}$, $b_c$ and $c_c$ are listed in Table~\ref{tab:cut_off}. 

\begin{figure}[tbp!]
\centering
\includegraphics[width=.4\textwidth]{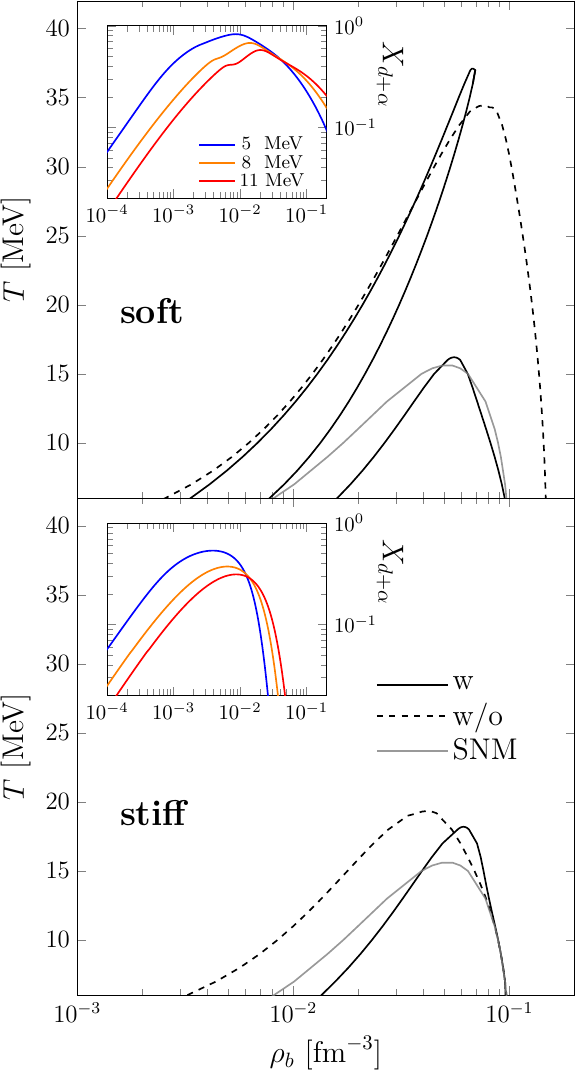}
\caption{Spinodal border in the ($\rho_{b}, T$) plane, as obtained for the two cutoff parametrizations
considered: soft (top) and stiff (bottom), by neglecting (dashed lines) or fully including (solid lines) in-medium effects. The spinodal boundary for pure nucleonic matter (gray) is shown for reference. The insets show the total cluster fraction $X_{d+\alpha}$ as function of $\rho_{b}$, for three representative temperature values.} 
\label{fig:spinodal_soft_vs_stiff}
\end{figure}
The total cluster mass fraction $X_{d+\alpha}$ correspondingly deduced as a function of the baryon density $\rho_{b}$ is shown in the insets of
Fig.~\ref{fig:spinodal_soft_vs_stiff} for three temperature values: $T = 5$, $8$ and $11$ MeV. As in Ref.~\cite{WangPRC2024}, for both soft (top panel) and stiff (bottom panel)
parametrizations, lower temperatures correspond to larger cluster
fractions below the Mott density, since the effective chemical
potentials decrease with $T$. Above the Mott density, the trend is
reversed: increasing the temperature reduces the effectiveness of
Pauli blocking, allowing clusters to persist up to higher densities.
Independently of the temperature, the soft parametrization predicts a
smoother suppression of clusters compared with the stiff one,
as shown in
Fig.~\ref{fig:X_c_soft_vs_stiff_d_and_a} at $T=8$ MeV. 

The main panels of
Fig.~\ref{fig:spinodal_soft_vs_stiff} display the corresponding spinodal borders in the
$(\rho_{b},T)$ plane for the two cutoff parametrizations: soft (top)
and stiff (bottom), obtained either by neglecting (dashed lines) or
fully including (solid lines) in-medium effects. When the density
dependence of the cutoff is neglected, the spinodal boundaries mainly
reflect the cluster fractions and their enhanced mean-field attraction,
leading to marked differences between the soft and stiff cases,
particularly at higher densities (and temperatures). In contrast, when in-medium effects are fully included, the additional attraction is largely compensated by a strong kinetic repulsion, bringing the spinodal border back close to that of pure nucleonic matter for both parametrizations.

However, in the soft case, a disjoint low-density instability region
emerges, which may extend to temperatures well above the critical
temperature of the liquid--gas phase transition 
of SNM~\cite{BorderiePPNP2019}. Its disappearance in the stiff case, however, highlights already its strong sensitivity to the details of the cutoff parametrization around the Mott density,  suggesting a likely
model dependence of this feature and motivating further investigation.
In the final part of this section, we therefore assess the sensitivity
of the spinodal border to additional ingredients of the calculation,
such as binding energies and the mean-field screening factor
$\eta$.

\singpart{Sensitivity to the binding energy and the mean-field screening factor}
\label{subsec:sensitivity_MF}

In the previous sections, we adopted the standard assumption that,
despite in-medium effects, nucleons bound in clusters experience the
same mean-field interaction as unbound nucleons and retain their vacuum masses. 

\begin{figure}[tbp!]
\centering
\includegraphics[width=.4\textwidth]{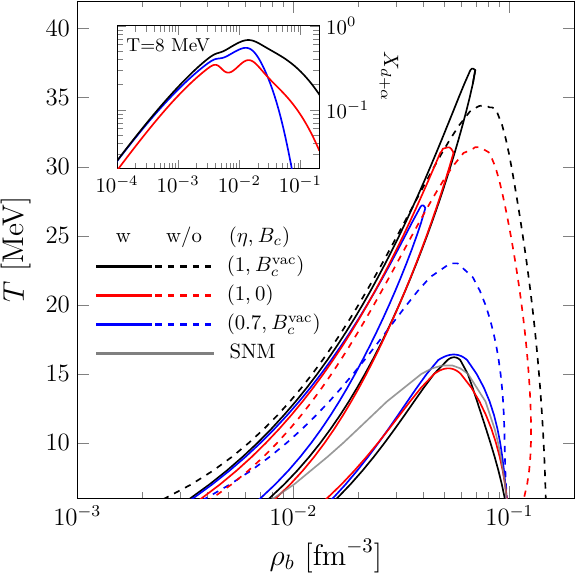}
\caption{Spinodal border in the $(\rho_{b}, T)$ plane obtained for
different values of the screening factor $\eta$ or the binding
energy $B_{c}$, by neglecting (dashed lines) or fully including (solid
lines) in-medium effects. The spinodal boundary for pure nucleonic
matter (gray) is shown for reference. The inset displays the density
dependence of the total cluster mass fraction $X_{d+\alpha}$ at
$T=8$ MeV.}
\label{fig:eta}
\end{figure}

A consistent implementation of density-, temperature-, and
momentum-dependent binding-energy shifts for both deuterons and
$\alpha$ particles—an approach first explored within a linear-response
framework in 
{Ref.~\cite{Pablo}} for nuclear matter with deuterons—would
go beyond the scope of the present work. Nevertheless, insight into the
role of the binding energy can be gained by considering the extreme
scenario in which clusters retain a vanishing binding energy,
$B_{c}=0$, instead of their vacuum value $B_{c}^{\rm vac}$.

Moreover, since the choice of the screening factor remains model
dependent and requires further microscopic
constraints~\cite{bougaultJPG2020,CustodioPRL2025}, we also explore the
sensitivity of our results to variations of $\eta$, in analogy with
Ref.~\cite{burEPJA2022}. 

In 
Fig.~\ref{fig:eta}, we show the spinodal boundary in the $(\rho_{b},T)$ plane for the soft cutoff parametrization, for which a disjoint low-density instability region emerges,  see the top panel of Fig.~\ref{fig:spinodal_soft_vs_stiff}. In addition to the standard choice
($\eta=1$, $B_{c}=B_{c}^{\rm vac}$) (black lines)  adopted in the analysis performed so far, 
we explore two cases: ($\eta=0.7$, $B_{c}=B_{c}^{\rm vac}$) (blue lines)  and
($\eta=1$, $B_{c}=0$) (red lines). The corresponding density behavior of the total cluster mass fractions $X_{d+\alpha}$ is shown in the inset, for the temperature $T = 8$ MeV. Both calculation schemes discussed above, i.e., neglecting (dashed lines) or
fully including (solid lines) the density dependence of the cutoff, are considered.

When neglecting 
in-medium effects, the spinodal boundary shrinks significantly as the screening factor $\eta$ is reduced. This behavior reflects the combined reduction of the cluster mass fraction
(see inset of Fig.~\ref{fig:eta}) and of the mean-field attraction
acting on bound nucleons. Consequently, the blue dashed curve approaches the spinodal boundary of pure nucleonic matter (gray curve).
A similar trend is observed when vanishing binding energies are
adopted, primarily due to the reduction of the cluster fraction while
the mean-field strength remains unchanged. The effect is more pronounced
for $\alpha$ particles, owing to their larger vacuum binding energy, 
as reflected in the stronger suppression of $X_{d+\alpha}$ at higher densities, where $\alpha$ particles dominate. However, the leading contribution 
to the quenching of instabilities 
stems from the mean-field screening: the shrinkage of the spinodal region is indeed more significant in the case with a scaled mean-field, even though the corresponding cluster fraction remains larger than in the vanishing-binding-energy scenario over a broad density interval. 

The trend is qualitatively modified once in-medium effects are
fully included. In that case, the spinodal boundary becomes much less
sensitive to both $\eta$ and $B_{c}$ 
over most of the phase diagram, while the additional low-density branch remains markedly dependent on these ingredients. 
This behavior further emphasizes the importance of
properly accounting for the density dependence of the cutoff and
highlights the model sensitivity of this disconnected low-density
instability region.

For brevity, we do not display the density dependence of $\Delta_{d}$ and $\Delta_{\alpha}$, since varying $\eta$ or $B_{c}$ neither qualitatively modifies the in- or out-of-phase pattern of cluster fluctuations relative to
nucleons nor induces significant quantitative changes. These findings confirm that the full calculations of the standard spinodal region are
robust not only against the choice of the specific cutoff parametrization
(see Fig.~\ref{fig:spinodal_soft_vs_stiff}), but also with respect to the residual model dependence
associated with the screening of the cluster mean-field interaction
and the omission of explicit mass-shift contributions.

\section{Conclusions}
\label{sec:conclusions}

In this work we have investigated the thermodynamic stability of
clusterized nuclear matter at sub-saturation densities, focusing on the
curvature properties of the free-energy density and the resulting convexity
conditions. Light clusters were included as explicit components of the
equilibrium composition, allowing us to analyze how their presence modifies
the stability domain and, in particular, the onset of the liquid--gas
spinodal region.

A central aspect of our study was the introduction of an
infrared cutoff in the momentum integrals entering the density and current moments. This cutoff provides an effective way of mimicking the suppression
of low-momentum quasiparticle states in a correlated medium. By deriving the
corresponding 
Euler equations 
we clarified the formal
consequences of such a modification: even when density independent, the
exclusion of soft modes alters the inertial properties of the fluid through a
renormalized effective mass; when the cutoff depends on the local densities,
a further term of rearrangement appears, 
which can be seen as an additional contribution to the standard thermodynamical pressure.
Moreover, because of the density-dependent cutoff, also chemical potentials acquire extra terms. 
These analytical results provide a
consistent theoretical basis for understanding how a modified phase-space
structure influences the stability properties of the composed system.

On the thermodynamic side, our {free-energy density} curvature analysis shows that light clusters
significantly modify the stability boundary of low-density matter. Their
formation can shift the location and
extension of the spinodal region.
{If the density dependence of the cutoff is
neglected 
in the stability analysis, 
the explicit inclusion of light clusters would enlarge the spinodal region, well beyond the empirical
evidences on the critical temperature of the liquid--gas phase transition in SNM.}
On the other hand, we show that when the infrared cutoff depends on the
density, 
the additional terms in the chemical
potentials
change the size of the spinodal region, making it compatible again with that of SNM. 
Moreover, the predicted spinodal boundary turns out to be remarkably robust against variations of the cutoff parametrization as well as the
additional ingredients of the calculation, such as the binding energies
and the mean-field screening factor, provided that the density
dependence of the cutoff is fully 
taken into account. 

The structure of the unstable modes inside the spinodal
region is also strongly affected by in-medium effects. Without 
accounting for 
the density
dependence of the cutoff 
in the stability analysis, 
clusters fluctuate in phase with nucleons,
enhancing the growth of density fluctuations. When in-medium effects
are consistently included, Pauli blocking introduces an effective
repulsion that may drive clusters out of phase with nucleons for a
sufficiently stiff cutoff, leading to a distillation-like mechanism, in which clusters are pushed toward low-density regions while nucleonic instabilities grow.
In the case of a  softer density dependence, however, clusters can survive at
moderate densities and cooperate with nucleons in the formation of
intermediate-mass fragments.
{The same features are observed within a dynamical description of the onset of instabilities, based either on the Euler equations or on the Vlasov approach.}

Since these instabilities determine the
propensity of nuclear matter to separate into dense and dilute phases, the
presence of clusters is expected to affect the early mechanisms that drive
fragmentation in heavy-ion collisions. 
{Possible scenarios, related to the density dependence of in-medium effects, range from the cooperation of clusters with nucleons, yielding the formation of larger fragments, to the occurrence of a distillation mechanism, where clusters populate the dilute phase and smaller fragments are formed.}
Our findings thus offer a framework to
interpret possible signatures of clustering in experimental observables
related to multifragmentation, isospin distributions, and fluctuations.

Moreover, the implications of this work extend beyond  
microphysics. Beyond terrestrial experiments, the same density and temperature ranges are relevant for supernova matter and the outer layers of neutron stars, where clusters and mechanical instabilities coexist and jointly shape the formation of nonuniform structures. 
A reliable description of dilute nuclear matter---including its clustering behaviour and dynamical instabilities---is crucial for modeling neutron-star
crusts and their response to density perturbations. These regions play a key
role in phenomena such as crust-breaking events, resonant crust--core oscillations, and the generation of gravitational-wave signals in binary
inspirals. As next-generation gravitational-wave observatories (such as the
Einstein Telescope and Cosmic Explorer) improve the precision with which the tidal response of neutron stars can be
measured, the low-density equation of state---including the effects discussed
here---will become increasingly important for accurate waveform modeling and
for extracting nuclear-physics information from astrophysical data.

Our results, linking microscopic quasiparticle dynamics to
macroscopic thermodynamic behavior in the presence of
momentum-space truncations, provide thus a theoretical framework for
applications to heavy ion collisions and neutron-star mergers, with the aim to assess the impact of these mechanisms on fragment formation and  astrophysical observables.

\begin{acknowledgments}
Stimulating discussions with Francesca Gulminelli, Pablo Nieto Gallego, Francesco Matera, Gerd R\"opke, and
Stefan Typel are gratefully acknowledged.
\end{acknowledgments}

\appendix
\begin{widetext}
\section{Chemical potentials and rearrangement terms}
\label{app:mu}

Thermodynamical consistency requires that the chemical potentials are
obtained, through Eq.~\eqref{eq:mu},
from the free-energy density functional $\mathcal{F}$ defined in Eq.~\eqref{eq:functional}. In the present
framework, $\mathcal{F}$ depends on the densities through:
(i) the potential energy density $\mathcal{U}(\{\rho_l\})$,
(ii) the effective chemical potentials $\mu_j^\ast$, and
(iii) the density-dependent infrared cutoffs $\lambda_c$.
Then, at fixed temperature, the total variation of the free energy reads
\begin{eqnarray}
\delta \mathcal{F}
& = & \delta \mathcal{U} + \sum_{jl} \rho_{j} \dfrac{\partial \mu_{l}^{\ast}}{\partial \rho_{l}} \delta \rho_{l} + \sum_{j} \left(m_{j} + \mu_{j}^{\ast} \right) \delta \rho_{j} - T \delta \left \{  \sum_{j} \dfrac{\alpha_{j}}{\sigma_{j}} \int_{\lambda_{j}}^{\infty} d\epsilon_{j} \sqrt{\epsilon_{j}} \ln \left[ 1 + \sigma_{j} \exp\left( - \dfrac{\epsilon_{j} - \mu_{j}^{\ast}}
{T}\right) \right] \right \} \nonumber \\
& = &  \sum_{j} \dfrac{\partial \mathcal{U}}{\partial \rho_{j}} \delta \rho_{j} + \sum_{jl} \rho_{j} \dfrac{\partial \mu_{j}^{\ast}}{\partial \rho_{l}} \delta \rho_{l} + \sum_{j} \left(m_{j} + \mu_{j}^{\ast} \right) \delta \rho_{j}  - \sum_{jl} \alpha_{j}\int_{\lambda_{j}}^{\infty} d\epsilon_{j} \sqrt{\epsilon_{j}} \left[ \exp\left( \dfrac{\epsilon_{j} - \mu_{j}^{\ast}}
{T}\right) + \sigma_{j}  \right]^{-1}  \dfrac{\partial \mu_{j}^{\ast}}{\partial \rho_{l}} \delta \rho_{l}  \nonumber \\
&& + T \sum_{c} \sigma_{c} \alpha_{c}
\sqrt{\lambda_{c}} \ln \left \{ \exp\left( - \dfrac{\lambda_{c} - \mu_{c}^{\ast}}
{T}\right) \left[\exp\left(  \dfrac{\lambda_{c} - \mu_{c}^{\ast}}
{T}\right) + \sigma_{c}  \right] \right \} \sum_{j} 
\dfrac{\partial \lambda_{c}}{\partial \rho_{ j}} \delta \rho_{j} \nonumber \\
& = &  \sum_{j} \dfrac{\partial \mathcal{U}}{\partial \rho_{j}} \delta \rho_{j} + \sum_{jl} \rho_{j} \dfrac{\partial \mu_{j}^{\ast}}{\partial \rho_{l}} \delta \rho_{l} + \sum_{j} \left(m_{j} + \mu_{j}^{\ast} \right)\delta \rho_{j}  - \sum_{jl} \dfrac{\partial \mu_{j}^{\ast}}{\partial \rho_{l}} \delta \rho_{l}   \alpha_{j}\int_{\lambda_{j}}^{\infty} d\epsilon_{j} \sqrt{\epsilon_{j}} f_{j} \nonumber \\ && - \sum_{j} \sum_{c} \sigma_{c} \alpha_{c} \sqrt{\lambda_{c}}\left( \lambda_{c} - \mu_{c}^{\ast} + T \ln f_{c}^{\lambda}\right) 
\dfrac{\partial \lambda_{c}}{\partial \rho_{ j}} \delta \rho_{j} \nonumber \\ \end{eqnarray}
having employed the definition of the equilibrium distribution
function $f_j$ [Eq.~\eqref{eq:distribution_function}] and introduced $\alpha_{j}$ [Eq.~\eqref{eq:alpha}] and  the shorthand notation
$f_c^\lambda \equiv f_c(\lambda_c)$.
Using also Eqs.~\eqref{eq:rhoi} and~\eqref{eq:potential},
the previous expression simplifies to
\begin{equation}
\delta \mathcal{F}
=
\sum_j
U_j \delta \rho_j
+
\sum_j
\left(m_{j} + \mu_{j}^{\ast} \right)\delta \rho_j
+
\sum_j 
\tilde{\mu}_{j}\, \delta \rho_j ,
\end{equation}
where
$\tilde{\mu}_{j}$ are additional rearrangement contributions
arising from the density dependence of the cutoff,
explicitly given by
Eq.~\eqref{eq:mutilde}, that is
\begin{equation}
\tilde{\mu}_{j}
=
-\sum_c \sigma_c \alpha_c \sqrt{\lambda_c}
\left(
\lambda_c - \mu_c^\ast
+ T \ln f_c^\lambda
\right)
\left.
\frac{\partial \lambda_c}{\partial \rho_j}
\right|_{\{\rho_l,\,l\neq j\}}.
\end{equation}

Finally, imposing the thermodynamical condition 
$\delta \mathcal{F} = \sum_j \mu_j \delta \rho_j$ [Eq.~\eqref{eq:mu1}],
one obtains
\begin{equation}
\mu_j
=
m_j
+
\mu_j^\ast
+
U_j
+
\tilde{\mu}_{j},
\end{equation}
which coincides with Eq.~\eqref{eq:mu}.

\section{Chemical potential derivatives and free-energy curvature matrix}
\label{app:dmu}

In this section, we evaluate the contributions entering the chemical
potential derivatives,
\begin{eqnarray}
\dfrac{\partial \mu_{j}}{\partial \rho_{l}}
=
\dfrac{\partial \mu_{j}^{\ast}}{\partial \rho_{l}}
+
\dfrac{\partial U_{j}}{\partial \rho_{l}}
+
\dfrac{\partial \tilde{\mu}_{j}}{\partial \rho_{l}} .
\label{eq:dmu}
\end{eqnarray}

The density derivatives of the mean-field potential $U_{j}$ are
straightforward to obtain. The derivatives of the effective chemical
potentials $\mu_j^\ast$ can instead be determined by observing,
in light of Eq.~\eqref{eq:rhoi}, that
\begin{eqnarray}
\delta_{jl}
&=&
\dfrac{\partial \rho_{j}}{\partial \rho_{l}}
=
\alpha_{j}
\dfrac{\partial}{\partial \rho_{l}}
\left[
\int_{\lambda_{j}}^{\infty}
d\epsilon_{j}\,
\sqrt{\epsilon_{j}}\, f_{j}
\right] = \alpha_{j}
\int_{\lambda_{j}}^{\infty}
d\epsilon_{j}\,
\sqrt{\epsilon_{j}}\,
\dfrac{\partial f_{j}}{\partial \mu_{j}^{\ast}}
\dfrac{\partial \mu_{j}^{\ast}}{\partial \rho_{l}}
-
\delta_{jc}\,
\alpha_{j}\sqrt{\lambda_{j}}\,
f_{j}^{\lambda}
\dfrac{\partial \lambda_{j}}{\partial \rho_{l}} ,
\end{eqnarray}
where $\lambda_{j}=0$ for neutrons and protons. Using the definition of $\Phi^{jl}$ given in Eq.~\eqref{eq:Phi}, one obtains
\begin{eqnarray}
\dfrac{\partial \mu_{j}^{\ast}}{\partial \rho_{l}}
&=&
\left(
\delta_{jl}
+
\delta_{jc}\Phi^{jl}
\right)
\dfrac{1}{\alpha_{j}}
\left[
\int_{\lambda_{j}}^{\infty}
d\epsilon_{j}\,
\sqrt{\epsilon_{j}}\,
\dfrac{\partial f_{j}}{\partial \mu_{j}^{\ast}}
\right]^{-1} = 
\left(
\delta_{jl}
+
\delta_{jc}\Phi^{jl}
\right)
N_{j}^{-1} ,
\label{eq:dmustar}
\end{eqnarray}
which is thus related, through Eq.~\eqref{eq:nj}, to the inverse
thermally averaged level density for any distribution function
depending on $(\epsilon_{j}-\mu_{j}^{\ast})$.

After straightforward algebra, the last contribution in
Eq.~\eqref{eq:dmu} reads
\begin{eqnarray}
\dfrac{\partial \tilde{\mu}_{j}}{\partial \rho_{l}}
&=&
-\sum_{c}
\Bigg[ \sigma_{c}
\alpha_{c}
\sqrt{\lambda_{c}}
\left(
\lambda_{c}-\mu_{c}^{\ast}
+
T\ln f_{c}^{\lambda}
\right)
\left(
\dfrac{1}{2\lambda_{c}}
\dfrac{\partial \lambda_{c}}{\partial \rho_{l}}
\dfrac{\partial \lambda_{c}}{\partial \rho_{j}}
+
\dfrac{\partial^{2}\lambda_{c}}
{\partial \rho_{l}\partial \rho_{j}}
\right)
+
\Phi^{cj}
\left(
\dfrac{\partial \lambda_{c}}{\partial \rho_{l}}
-
\dfrac{\partial \mu_{c}^{\ast}}{\partial \rho_{l}}
\right)
\Bigg] .
\label{eq:dmutilde}
\end{eqnarray}

Using Eq.~\eqref{eq:c3} together with
Eqs.~\eqref{eq:Phi}, \eqref{eq:landau}, \eqref{eq:nj}, and
\eqref{eq:phi_tilde}, it follows that, in the simplified case of a
single cluster species, the free-energy curvature matrix can be written as
\begin{equation}
\mathbb{C} =
\begin{pmatrix}
\dfrac{1 + F_{0} + 2\tilde{\Phi}^{qq}}{2N_{q}} &
\dfrac{F_{0}^{qd} + \tilde{\Phi}^{qd}}{N_{q}} \\[1ex]
\dfrac{F_{0}^{dq} + \tilde{\Phi}^{dq} + \Phi^{dq}}{N_{d}} &
\dfrac{1 + F_{0}^{dd} + \tilde{\Phi}^{dd} + \Phi^{dd}}{N_{d}}
\end{pmatrix} ,
\end{equation}
as given in Eq.~\eqref{eq:c2lambda}.

Finally, by combining Eqs.~\eqref{eq:dmustar} and~\eqref{eq:dmutilde}, one readily verifies that $\mathbb{C}_{12}=\mathbb{C}_{21}$, so that the curvature matrix $\mathbb{C}$ defined in Eq.~\eqref{eq:c2lambda} is symmetric, as required.

\section{Euler equations}

\label{app:euler}
Let us start from the collisionless Boltzmann equation, {namely the Vlasov equation,} for the distribution function $f_{j}$ (Eq.\ \eqref{eq:distribution_function}) of a species $j$ in phase space
\begin{equation}
\dfrac{\partial f_{j}}{\partial t} + \nabla_{\mathbf{r}} f_{j} \cdot \nabla_{\mathbf{p}} \varepsilon_{j} - \nabla_{\mathbf{p}} f_{j} \cdot \nabla_{\mathbf{r}} \varepsilon_{j} = 0  
\label{eq:vlasov}
\end{equation}
where 
\begin{equation}
\varepsilon_{j} = \epsilon_{j} + U_{j} + \tilde{\varepsilon}_{j}    
\end{equation}
is the single particle energy, $\epsilon_{j} = p^2/(2m_{j})$,  $U_{j}$ the mean-field potential and $\tilde{\varepsilon}_{j}$ the extra contribution arising from the density dependence of the cutoff (see Eqs.\  \eqref{eq:potential} and\  \eqref{eq:epsilonlambda}). 

The moments of the collisionless Boltzmann equation involve integrating the equation over momentum space, weighted by different functions of momentum, to obtain equations for macroscopic quantities like density, momentum, and energy. The lowest moments of the Boltzmann equation yield the equations of fluid dynamics, with the first moment (zeroth order in momentum) giving the continuity equation, the second moment (first order in momentum) giving the momentum equation (Navier-Stokes equation).

The Euler equations are a simplified form of the Navier-Stokes equations, neglecting viscosity and heat conduction.
The derivation of the Euler equations requires thus to combine the zeroth and first moment in momentum of Eq.\  \eqref{eq:vlasov},
{assuming an isotropic distribution function for the different species.}

\subsubsection{Zeroth moment}

The first contribution is
\begin{eqnarray}
g_{j} \int_{|\mathbf{p}|>\Lambda_{j}} \dfrac{d\mathbf{p}}{h^{3}} 
\dfrac{\partial f_{j}}{\partial t}
=
\dfrac{\partial}{\partial t}
g_{j} \int_{|\mathbf{p}|>\Lambda_{j}} 
\dfrac{d\mathbf{p}}{h^{3}} f_{j}
=
\dfrac{\partial \rho_{j}}{\partial t},
\end{eqnarray}
where $g_{j}$ is the spin degeneracy and the infrared cutoff $\Lambda_{j}$ is
included in the definition of the density (Eq.\ \eqref{eq:rhoi}).

The second contribution becomes
\begin{eqnarray}
g_{j} \int_{|\mathbf{p}|>\Lambda_{j}} \frac{d\mathbf{p}}{h^{3}}\,
\nabla_{\mathbf r} f_{j} \cdot \nabla_{\mathbf p} \varepsilon_{j}
&=&
\nabla_{\mathbf r} \cdot \left( g_{j} \int_{|\mathbf{p}|>\Lambda_{j}} \frac{d\mathbf{p}}{h^{3}}\, f_{j} \mathbf v_{j} \right) \;+\; \dfrac{g_{j}}{h^{3}} \int_{\Sigma_{\Lambda_{j}}}  d\Sigma_{\Lambda_{j}}\ f_{j}\, \mathbf v_{j} \cdot \nabla_{\mathbf r}\Lambda_{j} \nonumber\\ &=& \nabla_{\mathbf r} \cdot ( \rho_{j} \mathbf u_{j} )
\end{eqnarray}
where $\mathbf{v}_{j} = \dfrac{\mathbf{p}}{m_{j}}$ is the velocity and 
\begin{equation}
\mathbf{u}_{j} = \dfrac{1}{\rho_{j}} \left( g_{j} \int_{|\mathbf{p}|>\Lambda_{j}} \dfrac{d\mathbf{p}}{h^{3}} \ \ f_{j} \mathbf{v}_{j} \right) 
\label{eq:avg_velocity}
\end{equation}
is the average velocity.  
The surface term vanishes because the angular integral over $\Sigma_{\Lambda_{j}}$  of 
$\mathbf v_{j}$ over the sphere 
$|\mathbf p|=\Lambda_{j}$ is zero for an isotropic distribution 
$f_{j}$.

The third contribution is
\begin{eqnarray}
g_{j} \int_{|\mathbf{p}|>\Lambda_{j}} \dfrac{d\mathbf{p}}{h^{3}}
\,
\nabla_{\mathbf p} f_{j} \cdot \nabla_{\mathbf r}
\varepsilon_{j}.
\end{eqnarray}
Using integration by parts in momentum space,
\begin{eqnarray}
g_{j} \int_{|\mathbf{p}|>\Lambda_{j}} \dfrac{d\mathbf{p}}{h^{3}}
\,
\nabla_{\mathbf p} f_{j} \cdot \nabla_{\mathbf r}(U_{j}+\tilde{\varepsilon}_{j})
=
\dfrac{g_{j}}{h^{3}}
\int_{\Sigma_{\Lambda_{j}}} 
d\Sigma_{\Lambda_{j}}\,
f_{j}\,
\nabla_{\mathbf r}(U_{j}+\tilde{\varepsilon}_{j})
\cdot \hat{\mathbf n}.
\end{eqnarray}

Because $f_{j}$ is isotropic on the cutoff surface $|\mathbf p|=\Lambda_{j}$,
the angular integral of $\hat{\mathbf n}$ vanishes,
and therefore the whole contribution is zero.

\medskip

Collecting all terms, the zeroth moment yields the usual continuity equation,
\begin{equation}
\frac{\partial \rho_{j}}{\partial t}
+
\nabla_{\mathbf r}\!\cdot ( \rho_{j} \mathbf u_{j} )
= 0.
\end{equation}

\subsubsection{First moment}

The first contribution is
\begin{eqnarray}
g_{j} \int_{|\mathbf{p}|>\Lambda_{j}} \dfrac{d\mathbf{p}}{h^{3}} \mathbf{p} \dfrac{\partial f_{j}}{\partial t} = m_{j} \dfrac{\partial}{\partial t} \left[ g_{j} \int_{|\mathbf{p}|>\Lambda_{j}} \dfrac{d\mathbf{p}}{h^{3}} f_{j} \mathbf{v}_{j} \right] = m_{j} \dfrac{\partial}{\partial t} \left( \rho_{j} \mathbf{u}_{j} \right)
\end{eqnarray}

Let us try to write the second contribution
\begin{eqnarray}
g_{j} \int_{|\mathbf{p}|>\Lambda_{j}} \dfrac{d\mathbf{p}}{h^{3}} \, \mathbf{p} \, \nabla_{\mathbf{r}} f_{j} \cdot \nabla_{\mathbf{p}} \varepsilon_{j} & = & g_{j} \int_{|\mathbf{p}|>\Lambda_{j}} \dfrac{d\mathbf{p}}{h^{3}} \, \mathbf{p} \, \nabla_{\mathbf{r}} f_{j} \cdot \dfrac{\mathbf{p}}{m_{j}} 
\end{eqnarray}
in a more explicit way. Then for each cartesian component (for example $x$), one has
\begin{eqnarray}
g_{j} \int_{|\mathbf{p}|>\Lambda_{j}} \dfrac{d\mathbf{p}}{h^{3}} \ \ p_{x} \nabla_{\mathbf{r}} f_{j} \cdot \dfrac{\mathbf{p}}{m_{j}} & = & g_{j} \int_{|\mathbf{p}|>\Lambda_{j}} \dfrac{d\mathbf{p}}{h^{3}} \ \ p_{x} \left( \dfrac{\partial f_{j}}{\partial x} \dfrac{p_{x}}{m_{j}} + \dfrac{\partial f_{j}}{\partial y} \dfrac{p_{y}}{m_{j}} + \dfrac{\partial f_{j}}{\partial z} \dfrac{p_{z}}{m_{j}} \right) \nonumber \\
& = & g_{j} \int_{|\mathbf{p}|>\Lambda_{j}} \dfrac{d\mathbf{p}}{h^{3}} \ \ \dfrac{2}{3} \dfrac{p^{2}}{2 m_{j}} \dfrac{\partial f_{j}}{\partial x} \nonumber \\
& = & \dfrac{2}{3} \alpha_{j} \int_{\lambda_{j}}^{\infty} d\epsilon_{j} \left(\epsilon_{j}\right)^{3/2}\dfrac{\partial f_{j}}{\partial \mu_{j}^{\ast}} \sum_{l} \dfrac{\partial \mu_{j}^{\ast}}{\partial \rho_{l}} \dfrac{\partial \rho_{l}}{\partial x} 
\end{eqnarray}
Thus, in a vectorial form
\begin{eqnarray}
g_{j} \int_{|\mathbf{p}|>\Lambda_{j}} \dfrac{d\mathbf{p}}{h^{3}} \, \mathbf{p} \, \nabla_{\mathbf{r}} f_{j} \cdot \nabla_{\mathbf{p}} \varepsilon_{j} & = & - \dfrac{2}{3} \left( \sum_{l} \dfrac{\partial \mu_{j}^{\ast}}{\partial \rho_{l}} \nabla_{\mathbf{r}}\rho_{l} \right) \alpha_{j} \int_{\lambda_{j}}^{\infty} d\epsilon_{j} \left(\epsilon_{j}\right)^{3/2}\dfrac{\partial f_{j}}{\partial \epsilon_{j}}
\end{eqnarray}

Analogously, the third contribution is given by
\begin{eqnarray}   
- g_{j} \int_{|\mathbf{p}|>\Lambda_{j}} \dfrac{d\mathbf{p}}{h^{3}} \ \ \mathbf{p}\, \nabla_{\mathbf{p}} f_{j} \cdot \nabla_{\mathbf{r}} \varepsilon_{j} 
\end{eqnarray}
whose first component writes
\begin{eqnarray}   
- g_{j} \int_{|\mathbf{p}|>\Lambda_{j}} \dfrac{d\mathbf{p}}{h^{3}} \ \ p_{x}\, \nabla_{\mathbf{p}} f_{j} \cdot \nabla_{\mathbf{r}} \varepsilon_{j} & = & - g_{j} \int_{|\mathbf{p}|>\Lambda_{j}} \dfrac{d\mathbf{p}}{h^{3}} \ \ p_{x}\, \dfrac{\partial f_{j}}{\partial \epsilon_{j}} \nabla_{\mathbf{p}} \epsilon_{j} \cdot \nabla_{\mathbf{r}} \left( U_{j} + \tilde{\varepsilon}_{j}
\right) \nonumber \\
& = & - \left[ \sum_{l}  \dfrac{\partial}{\partial \rho_{l}} \left( U_{j} + \tilde{\varepsilon}_{j}
\right) \dfrac{\partial \rho_{l}}{\partial x} \right] \dfrac{2}{3} g_{j} \int_{|\mathbf{p}|>\Lambda_{j}} \dfrac{d\mathbf{p}}{h^{3}} \, \dfrac{\partial f_{j}}{\partial \epsilon_{j}} \dfrac{p^{2}}{2m_{j}}  
\end{eqnarray}
and, in a vectorial form,
\begin{eqnarray}   
- g_{j} \int_{|\mathbf{p}|>\Lambda_{j}} \dfrac{d\mathbf{p}}{h^{3}} \ \ p_{x}\, \nabla_{\mathbf{p}} f_{j} \cdot \nabla_{\mathbf{r}} \varepsilon_{j} & = & - \dfrac{2}{3} \left[ \sum_{l}  \dfrac{\partial}{\partial \rho_{l}} \left( U_{j} + \tilde{\varepsilon}_{j}
\right) \nabla_{\mathbf{r}} \rho_{l} \right]  \alpha_{j} \int_{\lambda_{j}}^{\infty} d\epsilon_{j} \left(\epsilon_{j}\right)^{3/2}\dfrac{\partial f_{j}}{\partial \epsilon_{j}}
\end{eqnarray}
The sum of the second and the third contribution is finally given by
\begin{eqnarray}   
g_{j} \int_
{\Lambda_{j}} 
\dfrac{d\mathbf{p}}{h^{3}} \, \mathbf{p} \left( \nabla_{\mathbf{r}} f_{j} \cdot \nabla_{\mathbf{p}} \varepsilon_{j} - \nabla_{\mathbf{p}} f_{j} \cdot \nabla_{\mathbf{r}} \varepsilon_{j} \right)  
& = & - \dfrac{2}{3} \left[ \sum_{l}  \dfrac{\partial}{\partial \rho_{l}}\left( \mu_{j}^{\ast} + U_{j} + \tilde{\varepsilon}_{j}\right) \nabla_{\mathbf{r}} \rho_{l} \right] \alpha_{j} \int_{\lambda_{j}}^{\infty} d\epsilon_{j} \left(\epsilon_{j}\right)^{3/2}\dfrac{\partial f_{j}}{\partial \epsilon_{j}}   \nonumber \\
& = & \left( \dfrac{2}{3}\alpha_{j} f_{j}^{\lambda} \lambda_{j}^{3/2} + \rho_{j}\right) \left[ \sum_{l} \left( \dfrac{\partial \mu_{j}}{\partial \rho_{l}} - \dfrac{\partial \left( \tilde{\mu}_{j} - \tilde{\varepsilon}_{j} \right) }{\partial \rho_{l}} \right) \nabla_{\mathbf{r}} \rho_{l} \right]   \nonumber \\
& = & 
\mathcal{Z}_{j}^{\lambda}  \left( \nabla_{\mathbf{r}} P + 
\nabla_{\mathbf{r}} \tilde{P}
\right) 
\label{eq:plambda}
\end{eqnarray}
with $\mathcal{Z}_{j}$ given by Eq.\ \eqref{eq:zeta} and 
the definition of $\tilde{P}$ introduced in analogy with the standard definition of the pressure $P$ given by Eq.\ \eqref{eq:pressure}.
So, finally, the Euler equations take the form
\begin{eqnarray}
\frac{\partial \rho_{j}}{\partial t}
+ \nabla_{\mathbf{r}} \cdot \left( \rho_{j} \mathbf{u}_{j} \right)
& = & 0 \nonumber \\
m_{j}^{\lambda} \frac{\partial}{\partial t} \left( \rho_{j} \mathbf{u}_{j} \right)
+ \nabla_{\mathbf{r}} P
+ \nabla_{\mathbf{r}} \tilde{P}
&=& 0 ,
\end{eqnarray}
with $m_{j}^{\lambda} = \dfrac{m_{j}}{\mathcal{Z}_{j}^{\lambda}}$, as in Eq.\ \eqref{eq:euler}.

\end{widetext}

\bibliographystyle{apsrev4-1}
\bibliography{ref}

\end{document}